\documentclass[journal]{IEEEtran}
\IEEEoverridecommandlockouts
\usepackage{cite}
\usepackage{amsmath,amssymb,amsfonts}
\usepackage{algorithmic}
\usepackage{graphicx}
\usepackage{textcomp}
\usepackage{xcolor}
\usepackage[utf8]{inputenc}
\usepackage{cite}
\usepackage{times}  
\usepackage{helvet} 
\usepackage{courier}  
\usepackage[hyphens]{url}  
\usepackage{graphicx}  
\usepackage{amsfonts}
\usepackage{amsmath}
\usepackage{amssymb}
\usepackage{bm} 
\newcommand{\vect}[1]{\boldsymbol{\mathbf{#1}}}
\usepackage{comment}
\usepackage[linesnumbered,ruled]{algorithm2e}

\usepackage{float}
\usepackage{csquotes}
\usepackage{amsthm}
\usepackage[linesnumbered,ruled]{algorithm2e}
\newtheorem{theorem}{\textbf{Theorem}}
\newtheorem{proposition}{\textbf{Proposition}}
\newtheorem{definition}{\textbf{Definition}}
\newtheorem*{proof*}{\textbf{Proof}}
\newtheorem{lemma}{\textbf{Lemma}}
\usepackage{multicol}
\usepackage[caption=false,font=footnotesize]{subfig}
\usepackage{float}
\usepackage{stfloats}
\usepackage[export]{adjustbox}
\usepackage{setspace}
\usepackage[switch]{lineno}

\def\BibTeX{{\rm B\kern-.05em{\sc i\kern-.025em b}\kern-.08em
    T\kern-.1667em\lower.7ex\hbox{E}\kern-.125emX}}
\begin{document}



\title{Data-Driven Random Access Optimization in Multi-Cell IoT Networks with NOMA}

\author{
    \IEEEauthorblockN{Sami Khairy\IEEEauthorrefmark{1}, Prasanna Balaprakash\IEEEauthorrefmark{2}, Lin X. Cai\IEEEauthorrefmark{1}, H. Vincent Poor\IEEEauthorrefmark{3}}\\
    \IEEEauthorblockA{\IEEEauthorrefmark{1}Department of Electrical and Computer Engineering, Illinois Institute of Technology, Illinois, USA }
    \IEEEauthorblockA{\IEEEauthorrefmark{2}Mathematics and Computer Science Division, Argonne National Laboratory, Illinois, USA}\\
    \IEEEauthorblockA{\IEEEauthorrefmark{3}Department of Electrical Engineering, Princeton University, Princeton, New Jersey, USA}
    \\
    \IEEEauthorblockA{Email: skhairy@hawk.iit.edu, pbalapra@anl.gov, lincai@iit.edu, poor@princeton.edu}
}

\maketitle
 
\begin{abstract}
Non-orthogonal multiple access (NOMA) is a key technology to enable massive machine type communications (mMTC) in $5$G networks and beyond. In this paper, NOMA is applied to improve the random access efficiency in high-density spatially-distributed multi-cell wireless IoT networks, where IoT devices contend for accessing the shared wireless channel using an adaptive $p$-persistent slotted Aloha protocol. To enable a capacity-optimal network, a novel formulation of random channel access management is proposed, in which the transmission probability of each IoT device is tuned to maximize the geometric mean of users' expected capacity. It is shown that the network optimization objective is high dimensional and mathematically intractable, yet it admits favourable mathematical properties that enable the design of efficient data-driven algorithmic solutions which do not require a priori knowledge of the channel model or network topology. A centralized model-based algorithm and a scalable distributed model-free algorithm, are proposed to optimally tune the transmission probabilities of IoT devices and attain the maximum capacity. The convergence of the proposed algorithms to the optimal solution is further established based on convex optimization and game-theoretic analysis. Extensive simulations demonstrate the merits of the novel formulation and the efficacy of the proposed algorithms.

\end{abstract}

\begin{IEEEkeywords}
Non-Orthogonal Multiple Access,  Random Access, Wireless IoT Networks, Machine Learning
\end{IEEEkeywords}

\section{Introduction}

A massive number of Internet-of-Things (IoT) devices will emerge in the market to enable advanced IoT based applications such as environmental monitoring, smart homes, smart transportation networks, and smart cities, to name a few. According to Cisco's most recent internet report \cite{Cisco}, IoT connections will reach $14.7$ billion by $2023$ which accounts for half of the global connected devices. Centralized and scheduling-based multiple access techniques cannot fully support such unprecedented growth in the number of IoT devices because scheduling the transmissions of a massive number of IoT devices would introduce significant computational and communication overheads. Due to their scalability and ease of implementation, distributed random-access-based wireless technologies such as Wi-Fi, Zigbee, and Aloha-based LoRaWAN, will have a major role in provisioning massive IoT access in beyond 5G systems  \cite{khairy2018renewal,khairy2019sustainable,el2018lora}. 

It is well recognized that Non Orthogonal Multiple Access (NOMA) can improve the channel access efficiency in 5G cellular networks by exploiting Successive Interference Cancellation (SIC) to decode non-orthogonal data transmissions 
\cite{ding2017application,ding2017survey,maraqa2019survey,ni2019analysis}. In slotted-Aloha systems, applying NOMA can significantly improve the channel access efficiency as users are randomly paired when they access the channel \cite{choi2017noma,choi2018game,choi2018multichannel,ziru}. Integrating NOMA with slotted-Aloha systems is therefore a promising solution to support massive machine type communications (mMTC) of IoT devices in beyond 5G networks \cite{khairy2020constrained}. Analyzing the performance of multi-cell slotted-Aloha systems with NOMA however, is very challenging due to the combinatorial space of possible transmissions and interference events that affect data decoding at the base stations. To circumvent this complexity, existing works mainly focused on a single-cell random access network where users can meet one of pre-specified target received power levels at the base stations to simplify the analysis \cite{choi2017noma,choi2018game, choi2018multichannel,ziru}. In a realistic large scale wireless IoT network, wireless users are spatially distributed in multiple cells with inter-cell interference, and perfect power control to meet the target received power levels at the base stations may not be available. Machine learning enables a data-driven approach for holistic system design, control, and optimization, which can be leveraged to study challenging multi-cell wireless IoT systems in 5G and beyond.

In this work, we consider a high-density spatially-distributed multi-cell wireless IoT network with a massive number of IoT devices. IoT devices send uplink sensor data to the base stations (BSs) using an adaptive  $p$-persistent slotted Aloha protocol. BSs, on the other hand, exploit power-domain SIC to decode concurrent transmissions of multiple users when possible, thus improving the random access efficiency. To attain a capacity-optimal massive IoT network, we study wireless channel access management of each individual IoT device to leverage the heterogeneity of users' channels. 
To the best of our knowledge, our work is the first work to study network performance of a high-density spatially-distributed multi-cell wireless IoT network with NOMA, and is the first to conceive a provably-optimal  data-driven framework for managing the random access probabilities of individual IoT devices in a  multi-cell network environment.

The main contributions of our work can be summarized as follows. First, we propose a novel formulation to manage channel access of individual IoT devices in a multi-cell wireless IoT network with NOMA. Specifically, we formulate the problem of channel access of IoT devices as a single stage 
optimization problem, where the objective is to maximize the geometric mean of users' expected capacity. Second, we show that this optimization objective, albeit being mathematically intractable and high dimensional, admits favourable mathematical properties which enable the design of efficient data-driven algorithmic solutions that avoid channel access starvation of users in multi-cell networks. Third, two learning based algorithms are proposed to optimally tune the transmission probabilities of IoT devices. The first algorithm is a centralized model-based algorithm in which a central controller learns an Input-Concave Neural Network (ICNN) to predict the system's performance. An upper confidence bound type approach is devised to strike a balance between model exploration and model exploitation; and the optimal transmission probabilities can be found by gradient ascent-based optimization on the learned model. The second algorithm is a scalable, distributed, model-free algorithm, in which each individual IoT device adapts its transmission probability locally based on an observable aggregate quantity that is a function of other users' transmission probabilities. Because the proposed algorithms are data-driven and enable continual learning in the sense that network capacity is improved as more algorithm iterations are executed, knowledge of the channel model or the network topology is not required. The convergence of the proposed algorithms to the optimal solution is further established based on convex optimization and game-theoretic analysis. Last but not least, we conduct extensive simulations to demonstrate the merits of  the novel formulation and the efficacy of the proposed algorithms. It is shown that optimizing the geometric mean of users' expected capacity
can greatly improve the throughput fairness among devices. 

The remainder of this paper is organized as follows. Related research works are given in Sec. II, followed by the system model and the problem formulation in Secs. III and  IV, respectively. The proposed centralized and distributed learning based algorithms are presented in Secs. V and VI. The performance evaluation results are provided in Sec. VII, followed by concluding remarks and future work in Sec. VIII.


\section{Related Works}


The application of NOMA in $5$G networks have been extensively studied in the literature. In \cite{ding2015impact,al2017optimum}, a centralized  scheduling-based network is considered, where the central controller schedules the transmissions of pairs of users, and decides their transmission powers according to their channel conditions, in order to maximize the sum-rate performance of NOMA users. Centralized scheduling, however, is not considered as a scalable solution for the unprecedented growth in the number of IoT devices. For instance, in mMTC $5$G based networks, scheduling transmissions of a massive number of IoT devices would incur significant overheads. In addition, it may be too expensive for low power IoT devices to estimate their channel conditions and provide the scheduler with channel state information in a timely manner. 

Recent works have proposed to apply NOMA in a slotted Aloha system such that users are randomly paired when they distributively access the channel to support mMTC in $5$G and beyond. Simulation results in \cite{elkourdi2018enabling} show that the throughput performance of slotted Aloha with NOMA significantly outperforms that of conventional slotted Aloha. A game-theoretic formulation to determine the transmission probability in a single-cell NOMA-based Aloha system is introduced in \cite{choi2018game,choi2018multichannel}. The proposed solution considers a payoff function based on an energy-efficiency metric, which ensures Nash equilibrium but not maximum throughput. In \cite{seo2018nonorthogonal,seo2018performance,seo2018two,seo2019uplink}, a slotted Aloha system with NOMA and random channel fading is investigated. In this single-cell system, users choose one of pre-specified target received power levels at the BS based on either their channel gain or the geographical region corresponding to the BS. The system is analyzed in terms of access delay, throughput, and energy efficiency, and it is was shown that the  achievable maximum throughput significantly outperforms that of conventional slotted Aloha without NOMA, which is consistent with the findings of \cite{elkourdi2018enabling}.

Single-cell multi-channel Aloha with NOMA is studied in \cite{choi2017noma} where different users with multiple transmission power levels are distributed in different channels. 
A closed-form expression for a lower-bound on the throughput was derived, and it was shown that applying NOMA can provide a higher throughput than multi-channel Aloha by exploiting power difference \cite{choi2017noma}. While increasing the number
of power levels in multi-channel slotted Aloha with NOMA leads to further gains in the maximum achievable throughput, the gains are slower than linear \cite{yu2020throughput}. Some works have also studied algorithm design in single-cell slotted Aloha systems with NOMA and two target received power levels in order to optimize network throughput in \cite{qu2019distributed,ziru}.


In all these aforementioned works, single-cell slotted Aloha wireless networks with NOMA are studied. Furthermore, efficient power control techniques are assumed such that users can meet the target received power levels at the BS. In practical networks, wireless IoT devices are likely distributed in a large scale multi-cell network with inter-cell interference and experience heterogeneous wireless fading channels, which make optimal power control very challenging if not impossible. To the best of our knowledge, our work is the first to propose a provably-optimal data-driven framework to optimize individual random channel access probabilities of IoT devices in a multi-cell network environment with NOMA. 

\section{System Model} \label{sec:sysmod}

We consider a multi-cell IoT network consisting of $M$ Base Stations (BSs) and $N$ IoT devices in which IoT devices transmit uplink sensor data to the BSs using an adaptive $p$-persistent slotted Aloha protocol. Let $\mathcal{M}=\{1,\cdots, M\}$ be the set of BSs, and $\mathcal{N}=\{1,\cdots, N\}$ be the set of IoT devices. Time is slotted into fixed-length communication slots indexed by $n$, that is, the $n$-th communication slot is $[t_n, t_{n+1})$, where $t_{n+1} - t_n = \Delta t,~\forall n$. 
In the beginning of a slot $n$, an IoT device, e.g., device $i \in \mathcal{N}$, attempts to access the channel with probability $p_i$ in order to transmit its most recent sensor data to the BS, which in turn relays the data to remote network servers. Channel access probability of individual devices, $p_i, \forall i \in \mathcal{N}$, is locally adapted in order to leverage the heterogeneity of users' channels and attain the maximum network capacity.

Each IoT device is served by the nearest BS. Let $\mathcal{N}_m \subseteq \mathcal{\mathcal{N}}$ denote the subset of IoT devices which are associated with BS $m$, such that $|\mathcal{N}_m|\leq N$, $\bigcup_{m=1}^M \mathcal{N}_m = \mathcal{N}$, $\mathcal{N}_i \cap \mathcal{N}_j= \phi, \forall i\neq j \in \mathcal{M}$. 
Let $u^{i,m}=\{0,1\}$ indicate whether IoT device $i$ associates with BS $m$. If BS $m$ is the closest to IoT device $i$, $u^{i,m}=1$, and $u^{i,l}=0, \forall l \neq m$. IoT devices transmit with a fixed transmission power of $P_{TX}$ watts. The power of the signal transmitted by an IoT device $i$ to BS $m$ in any given communication slot is typically subject to a random channel model which accounts for small-scale and large-scale fading. In this work, we study the design of a data-driven network optimization framework which only depends on observing an aggregate quantity of users' throughputs as a feedback mechanism, and hence knowledge of the channel model or network topology are not required. Let $P^{i,m}_{RX}$ denote the received power at BS $m$ from IoT device $i$ in a communication slot $t_n$ given that user $i$ transmits. BS $m$ first attempts to decode the signal with the highest signal power under the interference from all other IoT devices involved in the NOMA transmissions. Without loss of generality, IoT devices which transmit in slot $n$ are sorted in the descending order of the received signal strength at BS $m$, such that $i=1$ is the IoT device with the highest received signal to interference plus noise (SNIR) at BS $m$, and $i=2$ is the IoT device with the second highest received SNIR at BS $m$ \footnote{Due to the decoding complexity of SIC, we consider that the two highest received signals of NOMA transmissions are possibly decodable, yet our proposed framework is readily extensible for more than 2 NOMA transmissions.}. Because of the distributed random channel access protocol, users $i\in\{1,2\}$ are randomly paired when they distributively access the channel. Let $\mathcal{I}_m$ be the random set of transmitters in slot $t_n$. The highest received SNIR at BS $m$ in slot $t_{n}$ is therefore,  
\begin{equation}
    \text{SNIR}_{1,m} = \frac{P^{1,m}_{RX}}{n_0 + \sum_{k \in \mathcal{I}_m \setminus \{1\}} P^{k,m}_{RX}},
\end{equation}
where $n_0$ is the noise floor power. Similarly, the second highest received SNIR at BS $m$ in slot $t_{n}$ is,
\begin{equation}
    \text{SNIR}_{2,m} = \frac{P^{2,m}_{RX}}{n_0 + \sum_{k \in \mathcal{I}_m\setminus \{1,2\} } P^{k,m}_{RX}},
\end{equation}
BS $m$ can decode the signal with $\text{SNIR}_{1,m}$ if 
\begin{enumerate}
    \item user 1 is associated with BS $m$, $u_n^{1,m}=1$, and,
    \item $\text{SNIR}_{1,m}$ is larger than or equal to the SNIR threshold $\text{SNIR}_\text{Th}$, i.e.,  $\Phi(\text{SNIR}_{1,m})=\text{SNIR}_{1,m}$,
\end{enumerate}
where $\Phi(.)$ is a threshold function to maintain a minimum target SNIR and quality of service, that is, $\Phi(\text{SNIR}_{i,m})=\text{SNIR}_{i,m}$ if $\text{SNIR}_{i,m} \geq \text{SNIR}_\text{Th}$, and $\Phi(\text{SNIR}_{i,m})=0$ otherwise. 
In addition, BS $m$ can decode the signal with $\text{SNIR}_{2,m}$ if
\begin{enumerate}
    \item The signal with $\text{SNIR}_{1,m}$ is successfully decoded,
    \item user 2 is associated with BS $m$, $u_n^{2,m}=1$, and,
    \item $\text{SNIR}_{2,m}$ is larger than or equal to the SNIR threshold, i.e., $\Phi(\text{SNIR}_{2,m})=\text{SNIR}_{2,m}$
\end{enumerate}
The upper bound on the rate achieved by the $i$-th user in slot $t_n$ given that user $i$ transmits is,
\begin{equation}
\begin{aligned}
     R_{i,\text{TX}} =  &\mathcal{W}\text{log}_2\Big(1+\Phi(\text{SNIR}_{1,m})u_n^{1,m}\Big)\mathbb{I}_{1,i} +\\
    &\mathcal{W}\text{log}_2\Big(1+\Phi(\text{SNIR}_{2,m})u_n^{1,m}u_n^{2,m}e_n^{1,m}\Big)\mathbb{I}_{2,i} 
\end{aligned}
\end{equation}
where $\mathcal{W}$ is the transmission bandwidth, $e_n^{1,m} = 1$ if  $\Phi(\text{SNIR}_{1,m})=\text{SNIR}_{1,m}$ and $0$ otherwise, $\mathbb{I}_{1,i}=1$ if user $i$ is the IoT device with the highest received SNIR at BS $m$ and $0$ otherwise, and $\mathbb{I}_{2,i}=1$ if user $i$ is the IoT device with the second highest received SNIR at BS $m$ and $0$ otherwise. Notice that either $\mathbb{I}_{1,i}$ or  $\mathbb{I}_{2,i}$ can be $1$ in slot $t_n$. It is worth mentioning that the maximum achievable rate by the $i$-th user in slot $t_n$ depends on the transmission probability vector of all IoT devices, $\vect{p} = (p_1, \cdots, p_N)$, and has the  general form,
\begin{equation} 
\begin{aligned} 
\label{form}
    R_i(\vect{p}) = &A_i \Big( p_i\prod_{j\neq i}(1-p_j) \Big) + \sum_{j\neq i} B^j_i \Big( p_i p_j\prod_{k\neq j,i}(1-p_k) \Big) + \\
    &\sum_{j \neq i} \sum_{k \neq i,j} C^{j,k}_i \Big( p_i p_j p_k \prod_{l\neq j,i,k}(1-p_l) \Big) + \cdots
\end{aligned}
\end{equation}
where  $A_i$ is a non-negative random variable (R.V.) which represents user $i$'s maximum theoretical rate given that user $i$ is the only user that transmits in slot $t_n$, $B^j_i$ is a non-negative R.V. which represents user $i$'s maximum theoretical rate given that user $i$ and another user $j$ transmit concurrently in $t_n$, and $C^{j,k}_i$ is a non-negative R.V. which represents user $i$'s {maximum theoretical} rate given that user $i$ and two other users, $j,k$, transmit concurrently in $t_n$, and so on and so forth.  Notice that $R_i(\vect{p})$ is a weighted sum of $\sum_{k=0}^{N-1} \binom{N-1}{k}= 2^{N-1}$ random variables. In theory, the distributions of $A_i,B^j_i,C^{j,k}_i,\cdots$ or their expectations can be derived by considering NOMA decoding events at the BSs, which {require knowledge} of the network topology and the random channel model. However, this is a very challenging task {in practice because of the massive number of IoT devices and the combinatorial number of possible transmission events}. In this paper, we propose a novel {data-driven formulation that enables the design of efficient algorithms to tune $\vect{p}$ so that the capacity of the network is optimized, without assuming knowledge of the network topology or the channel model.}    

\section{Problem Formulation}

To {maximize NOMA's gain by leveraging the heterogeneity of users' channels, and} enable a capacity-optimal network, we propose a novel formulation to manage random channel access of {individual} IoT devices in a high-density spatially-distributed multi-cell wireless IoT network. Specifically, we formulate the decision problem of tuning $p_i \in (0,1], \forall i \in \mathcal{N}$, as a single stage 
optimization problem, where the optimization objective, $\mathcal{O}(\vect{p})$, is the log of the geometric mean of users’ expected rates,
\begin{equation}\label{eq:system}
\begin{aligned}
\mathcal{O}(\vect{p}) &= \text{log} \Bigg[ \Big(\prod_{i=1}^N \Bar{R_i}(\vect{p}) \Big)^{1/N} \Bigg]\\
&= \frac{1}{N} \sum_{i=1}^N \text{log}\big(\Bar{R_i}(\vect{p}) \big) 
\end{aligned}
\end{equation}
where $\Bar{R_i}(\vect{p}) = \mathbb{E}[R_i(\vect{p})] \in \mathbb{R}_+, \forall i \in \mathcal{N}$. 
This formulation is motivated by the following merits of the geometric mean,
\begin{enumerate}
    \item By maximizing the log of the geometric mean of users' expected rates, the geometric mean of users' expected rates is maximized as the log function is a monotonic transformation that preserves the locations of maxima. Based on the arithmetic mean (AM) and geometric mean (GM) inequality, the geometric mean of user's expected rates is upper bounded by the arithmetic mean,
    $$\Big(\prod_{i=1}^N \Bar{R_i}(\vect{p}) \Big)^{1/N} \leq \frac{1}{N} \sum_{i=1}^N \Bar{R_i}(\vect{p}),$$
    and the equality holds when all users have the same expected rate, i.e., $\Bar{R_i}(\vect{p})=\Bar{R}(\vect{p}), ~\forall i$. Hence, by maximizing the geometric mean of users' expected rates, a lower bound on the average of users' expected rates is maximized.
    \item The geometric mean is a non-decreasing monotone function of users' expected rates: by increasing the  expected rate of any user, the geometric mean increases.
    \item The geometric mean is not as sensitive to outliers or extreme values as the arithmetic mean, and it is $0$ if any user has a zero expected rate. 
    Therefore, by maximizing the geometric mean {as a function of the high-dimensional transmission probability vector $\vect{p} \in (0,1]^N$}, we ensure that no user is starved under the optimal transmission probability vector $\vect{p}^*$, i.e.,  $\Bar{R_i}(\vect{p^*}) > 0, \forall i \in \mathcal{N}$. This is not achieved when the arithmetic mean is optimized {as a function of $\vect{p}$} \footnote{Notice that the AM of expected user rates is a convex combination of the achievable conditional expected rates by $2^N-1$ different transmissions events, and hence the maximum is attained at one of the extreme points.}, because the optimal transmission probability vector $\vect{p}^{AM}$ which maximizes the arithmetic mean grants the channel to users $\mathcal{N}^{AM}$ whose transmissions maximize the NOMA sum-rate performance, i.e., $p_i^{AM}=1, i\in \mathcal{N}^{AM}$, while other users $j \not \in \mathcal{N}^{AM}$ are starved $p_j^{AM}=0$, leading to very poor rate fairness. 
    Our proposed geometric-mean-based formulation on the other hand, natively ensures good rate fairness without imposing explicit fairness constraints, as will be shown in our extensive numerical experiments. 
\end{enumerate}
In addition, the optimization objective in \eqref{eq:system} admits attractive mathematical properties which facilitate the design of efficient centralized and provably convergent distributed optimization algorithms. Given some topology dependent conditions, we prove that $\mathcal{O}(\vect{p})$ is a strictly concave function on a convex set $S^{N}$ for $N\in \{2,3\}$. The proof of a general case of $N > 3$ is mathematically intractable, yet we conjecture that this property holds based on our extensive numerical experiments. In the following section, we show how this property can be used to design an efficient centralized learning-based algorithm to maximize \eqref{eq:system}. In Sec. VI, we propose a distributed learning-based algorithm to maximize \eqref{eq:system} and prove its convergence to a pure Nash equilibrium for the general case of $N$ users. 

To prove that $\mathcal{O}(\vect{p})$ is a strictly concave function on $S^{N}$ for the case of $N \in \{2,3\}$ users, we first prove that $\bar{R}_i(\vect{p})$ is a strictly log-concave function on a convex set $S^{N}_i$.
\begin{definition}
A non-negative function $f: \mathbb{R}^n \rightarrow \mathbb{R}_+$ is a strictly logarithmically concave (log-concave) if its domain (dom) is a convex set, and if it satisfies the inequality
$$ f(\beta \vect{x}+(1-\beta)\vect{y}) > f(\vect{x})^\beta f(\vect{y})^{1-\beta}$$ 
$\forall \vect{x},\vect{y} \in$ dom $f$ and $0 < \beta < 1$. If $f$ is a strictly positive function, $f(\vect{x})>0, \forall \vect{x} \in$ dom $f$, then $f$ is a strictly log-concave function if, 
$$ \text{log}(f(\beta\vect{x}+(1-\beta)\vect{y})) > \beta \text{log}(f(\vect{x})) + (1-\beta)\text{log}(f(\vect{y}))$$
in other words, $\text{log}(f)$ is a strictly concave function. 
\end{definition}
Furthermore, let $\mathbb{H}$ denote the hessian operator, 
\begin{equation}
    \mathbb{H}[f(\vect{p})] = \begin{bmatrix}
    \frac{\partial f  }{\partial p_1 \partial p_1} & \cdots & \frac{\partial f}{\partial p_1 \partial p_N}\\
    \vdots & \ddots & \vdots \\
    \frac{\partial  f}{\partial p_N\partial p_1}& \cdots & \frac{\partial  f}{\partial p_N \partial p_N}
    \end{bmatrix}.
\end{equation}
\begin{theorem}
In a multi-cell $p$-persistent slotted Aloha system of $N=2$ users and $M\in \{1,2\}$ BSs with NOMA, where users attempt to transmit to the closest BS with probability $p_i, \forall i \in \{1,2\}$ in each slot, the expected rate of a user $i$, $\bar{R}_i(p_1,p_2), \forall i \in \{1,2\}$, is a strictly log-concave function of $(p_1,p_2)$ on the convex set $S^2_i =\{(p_1,p_2) |~p_1,p_2 \in (0,1]\}, \forall i \in \{1,2\}$.
\end{theorem}
\begin{proof*}
A sufficient condition for 
$\text{log}\big(\bar{R}_i({\vect{p}})\big), \forall i \in \{1,2\}$, to be strictly concave is that $\mathbb{H}\Big[\text{log}\big(\bar{R}_i(\vect{p})\big)\Big] \prec 0$, i.e., the hessian matrix is negative definite $\forall \vect{p} \in S^N_i$. A necessary and sufficient condition for negative definiteness of the hessian matrix $\forall \vect{p} \in S^N_i$ is 
$(-1)^k D_k > 0$ for $k = 1, ..., N$ and $\forall \vect{p} \in S^N_i$, where $D_k$ is the determinant of the $k$-th leading principal sub-matrix of the hessian. Without loss of generality (W.L.O.G.), set $i=1$. The log of the expected rate of user $1$ is,
\begin{equation}
\text{log}\Big(\bar{R}_1(p_1,p_2)\Big) = \text{log}\Big(\mathbb{E}[A_1] p_1(1-p_2)  + \mathbb{E}[B^2_1] p_1 p_2\Big),
\end{equation}
with a hessian matrix,
\begin{equation}
    \mathbb{H}\Big[\text{log}\Big(\bar{R}_1(p_1,p_2)\Big)\Big] = \begin{bmatrix}
    \frac{-1}{p_1^2} & 0\\
    0 & \frac{-(\mathbb{E}[A_1]-\mathbb{E}[B^2_1])^2}{((1-p_2)\mathbb{E}[A_1]+\mathbb{E}[B^2_1])p_2)^2}.
    \end{bmatrix}
\end{equation}
As $(-1)^1D_1 = \frac{1}{p_1^2} > 0$ and $(-1)^2D_2 = \frac{(\mathbb{E}[A_1]-\mathbb{E}[B^2_1])^2}{p_1^2((1-p_2)\mathbb{E}[A_1]+\mathbb{E}[B^2_1])p_2)^2}   > 0 $, $\text{log}\Big(\bar{R}_1(p_1,p_2)\Big)$ is strictly concave on 
$S^2_1$. Similarity, $\text{log}\Big(\bar{R}_2(p_1,p_2)\Big)$ is strictly concave on $S_2^2$. $\blacksquare$
\end{proof*}

For $N=2$ case, $\text{log}\Big(\bar{R}_i(p_1,p_2)\Big)$ is strictly log-concave on $S_i^2$, and independent of $\mathbb{E}[A_i], \mathbb{E}[B_i^j]$, i.e., the property is independent of the topology and the achievable  expected user rate with and without interference from the other user. For $N=3$ case, $\text{log}\Big(\bar{R}_i(p_1,p_2,p_3)\Big)$ is strictly log-concave over a convex set $S^3_i, \forall i \in \{1,2,3\}$, if $\mathcal{C}_i= \bar{A}_i - \bar{B}_i^j - \bar{B}_i^k + \bar{C}_i^{j,k} > 0, \forall i,j,k \in \{1,2,3\}$ and  $i\neq j \neq k$, where $\bar{A}_i=\mathbb{E}[{A}_i], \bar{B}_i^j=\mathbb{E}[{B}_i^j]$, and $\bar{C}_i^{j,k}= \mathbb{E}[{C}_i^{j,k}]$, as formalized below.
\begin{theorem}
In a multi-cell $p$-persistent slotted Aloha system of $N=3$ users and $M\in\{1,2,3\}$ BSs with NOMA, where users attempt to transmit to the closest BS with probability $p_i, \forall i \in \{1,2,3\}$ in each slot, the expected rate of a user $i$, $\bar{R}_i(p_1,p_2,p_3), \forall i \in \{1,2,3\}$, is a strictly log-concave function of $(p_i,p_j,p_k)$ on the convex set $S_i^3 = \{(p_i,p_j,p_k)|~p_i \in (0,1],~p_j \in (0, \Tilde{p}_j),~p_k \in (0, \Tilde{p}_k)\}$, $\forall i,j,k\in\{1,2,3\}$ and $i \neq j \neq k$, where $\Tilde{p}_j = \frac{\bar{A}_i - \bar{B}_i^k - \sqrt{ \left\lvert \bar{A}_i\bar{C}_i^{j,k} - \bar{B}_i^j\bar{B}_i^k \right\rvert  }}{\mathcal{C}_i}$, and $\Tilde{p}_k = \frac{\bar{A}_i - \bar{B}_i^j - \sqrt{\left\lvert\bar{A}_i\bar{C}_i^{j,k} - \bar{B}_i^j\bar{B}_i^k \right\rvert}}{\mathcal{C}_i}$.
\end{theorem}
\begin{proof*}
W.L.O.G., let $i=1,j=2,$ and $k=3$. The log of the expected rate of user $1$ is,
\begin{equation}
\resizebox{1\hsize}{!}{$
\begin{aligned}
\text{log}\Big(\bar{R}_1(&p_1,p_2,p_3)\Big) =
\text{log}\Big(\bar{A}_1 p_1(1-p_2)(1-p_3) \\
&+ \bar{B}^2_1 p_1 p_2(1-p_3) + \bar{B}^3_1 p_1 p_3(1-p_2) + \bar{C}_1^{2,3}p_1p_2p_3\Big),
\end{aligned} $}
\end{equation}
with a hessian matrix whose entries are given by $\mathbb{H}_{1,1}= \frac{-1}{p_1^2}$, $\mathbb{H}_{1,2}=0$, $\mathbb{H}_{1,3}=0$, $\mathbb{H}_{2,1}=0$,
$\mathbb{H}_{2,2} = \frac{-\big[\bar{A}_1(1-p_3) - \bar{B}^2_1(1-p_3) + (\bar{B}^3_1-\bar{C}^{2,3}_1)p_3 \big]^2}{\big[\mathcal{F}(p_2,p_3) \big]^2}$, $\mathbb{H}_{2,3}= \mathbb{H}_{3,2}= \frac{\bar{A}_1\bar{C}^{2,3}_1 - \bar{B}^2_1\bar{B}^3_1}{\big[\mathcal{F}(p_2,p_3) \big]^2}$,
$\mathbb{H}_{3,1}=0$, and $\mathbb{H}_{3,3} = \frac{-\big[\bar{A}_1(1-p_2) - \bar{B}^3_1(1-p_2) + (\bar{B}^2_1-\bar{C}^{2,3}_1)p_2 \big]^2}{\big[\mathcal{F}(p_2,p_3) \big]^2}$, where $\mathcal{F}(p_2,p_3)= \big(\bar{A}_1 (1-p_2)(1-p_3)+ \bar{B}^2_1 p_2(1-p_3) + \bar{B}^3_1  p_3(1-p_2) + \bar{C}_1^{2,3}p_2p_3\big)$. A sufficient condition for  $\text{log}\Big(\bar{R}_1(p_1,p_2,p_3)\Big)$ to be strictly concave on $S_1^3$ is that $\mathbb{H} \prec 0, \forall \vect{p}\in S_1^3$, which is true if and only if the leading principle minors satisfy: (C1) $(-1)^1D_1 = -\mathbb{H}_{1,1} > 0$, (C2) $(-1)^2D_2 = \mathbb{H}_{1,1}\mathbb{H}_{2,2} > 0$, and (C3) $(-1)^3D_3 = -\mathbb{H}_{1,1}\big(\mathbb{H}_{2,2}\mathbb{H}_{3,3} - \mathbb{H}_{2,3}^2 \big) > 0$, $\forall \vect{p}\in S_1^3$. It can be observed that (C1) and (C2) are always satisfied, so it remains to investigate (C3). Notice that $-\mathbb{H}_{2,2}>0, -\mathbb{H}_{3,3} >0$, and suppose that (C3-1) $-\mathbb{H}_{2,2} > |\mathbb{H}_{2,3}|$ and (C3-2) $-\mathbb{H}_{3,2} > |\mathbb{H}_{2,3}|$ hold. Taking the log of (C3-1) and (C3-2) and adding the resulting two inequalities we obtain $\text{log}(-\mathbb{H}_{2,2}) + \text{log}(-\mathbb{H}_{3,3}) > 2\text{log}\big(|\mathbb{H}_{2,3}|\big)$, which can be simplified to $\text{log}(\mathbb{H}_{2,2}\mathbb{H}_{3,3}) > \text{log}\big(\mathbb{H}_{2,3}^2\big)$. By exponentiating both sides and rearranging the terms, we have that  $\big(\mathbb{H}_{2,2}\mathbb{H}_{3,3}-\mathbb{H}_{2,3}^2 \big)> 0$. Since $-\mathbb{H}_{1,1} > 0$, we can conclude that (C3) is satisfied if the alternate conditions (C3-1) and (C3-2) are both satisfied. Hence, we now show that (C3-1) is satisfied $\forall p_3 \in (0,\Tilde{p}_3)$. Notice that (C3-1) is satisfied if $\big[\bar{A}_1(1-p_3) - \bar{B}^2_1(1-p_3) + (\bar{B}^3_1-\bar{C}^{2,3}_1)p_3 \big]^2 - |\bar{A}_1\bar{C}^{2,3}_1 - \bar{B}^2_1\bar{B}^3_1| = \big(\mathcal{C}_1\big)^2p_3^2 - 2\mathcal{C}_1(\bar{A}_1-\bar{B}_1^2)p_3 + (\bar{A}_1-\bar{B}_1^2)^2 - |\bar{A}_1\bar{C}^{2,3}_1 - \bar{B}^2_1\bar{B}^3_1| > 0 $, which is true  $\forall p_3 < \Tilde{p}_3$ as $\Tilde{p}_3$ is the smaller root of the convex quadratic equation in $p_3$. Similarly, it can also be shown that (C3-2) is satisfied $\forall p_2 < \Tilde{p}_2$.  $\blacksquare$
\end{proof*}

Because the non-negative sum of concave functions is also concave, we can conclude based on the result of theorem 1 that $\mathcal{O}(p_1,p_2)$ is a strictly concave function on the convex set $S^2=S_1^2\cap S_2^2$ when $N=2$. For $N=3$, $\mathcal{O}(p_1,p_2,p_3)$ is a strictly concave function on the convex set $S^3=S_1^3\cap S_2^3\cap S_3^3$ if $\mathcal{C}_i= \bar{A}_i - \bar{B}_i^j - \bar{B}_i^k + \bar{C}_i^{j,k} > 0, \forall i,j,k \in \{1,2,3\}$ and  $i\neq j \neq k$. It is worth mentioning that the condition $\mathcal{C}_i > 0$ means that the interference caused by users $j,k$ to transmission of user $i$ is non-negligible and degrades the achievable expected rate of user $i$, such that the surface of $\text{log}\Big(\bar{R}_i(p_i,p_j,p_k)\Big)$ has no flat regions w.r.t. $p_j$ or $p_k$. To further elaborate, consider a simple case of a large-scale fading channel. $\mathcal{C}_i> 0$ becomes ill-conditioned in the case where user $j$, or user $k$, or both, are far away from BS $m$ of user $i$ $(u^{i,m}=1)$, to the extent that their received transmission power at BS $m$ is much smaller than the order of the noise power $n_0$. As a specific example, 
suppose that $d_{j,m} >> d_{i,m}$  while $d_{k,m}$ is on the same order of $d_{i,m}$. In such case, we have $\bar{B}_i^j \rightarrow \bar{A}_i, \bar{C}_i^{j,k} \rightarrow \bar{B}_i^k$, then $\text{log}\Big(\bar{R}_i(p_i,p_j,p_k)\Big)\rightarrow
\text{log}\Big(\bar{A}_i p_i(1-p_k)+ \bar{B}^k_i p_i p_k\Big)$. This implies that, i) $\frac{\partial }{\partial p_j} \text{log}\Big(\bar{R}_i(p_i,p_j,p_k)\Big) = 0$, i.e., user $j$'s transmission strategy has no impact on user $i$, and $p_j$ can be set to any arbitrary value $p_j \in (0,1]$ without impacting $\text{log}\Big(\bar{R}_i(p_i,p_j,p_k)\Big)$; ii) for any $p_j \in (0,1]$, $\text{log}\Big(\bar{R}_i(p_i,p_j,p_k)\Big)$ is a strictly concave function of $p_i,p_k$ on the convex set $S^2_i$ based on the result of theorem 1; and iii) $\mathcal{O}(p_1,p_2,p_3)$ is a strictly concave function on the convex set $S^3=\Tilde{S}_i\cap S_j^3\cap S_k^3=S_j^3\cap S_k^3$, if $\mathcal{C}_j > 0$ and  $\mathcal{C}_k > 0$, where $\Tilde{S}_i =\{(p_1,p_2,p_3) |~p_1,p_2,p_3 \in (0,1]\}$.

For a general case of $N > 3$ users, we conjecture that $\mathcal{O}(\vect{p})$ is a concave function over a convex set $S^N$, such that convex programming techniques can be applied to find the globally optimal transmission probability vector $\vect{p}^*$, although $\mathcal{O}(\vect{p})$ is not readily available in a closed-form. To obtain a closed form expression of $\mathcal{O}(\vect{p})$, the expectations of $N*\sum_{k=0}^{N-1} \binom{N-1}{k}= N*2^{N-1}$ R.V.s should be derived by considering each of $2^N-1$ possible transmission events, and evaluating the expectations based on network topology and the random channel model. This is not practically feasible for high-density networks with a large $N$. Estimating these expectations via Monte Carlo simulations can also be computationally prohibitive. Machine learning on the other hand, provides a data driven approach for system design, optimization, and control, and can be exploited to maximize \eqref{eq:system}. In the following section, a centralized learning-based optimization algorithm is proposed to optimize \eqref{eq:system}.


\section{Centralized Learning-Based Optimization Algorithm}

In this section, we design an active-learning-based algorithm in which a compatible surrogate model for $\mathcal{O}(\vect{p})$ is sequentially learned and optimized based on online-sampled training data, to find the optimal transmission probability vector $\vect{p}^*$. Notice that for a given transmission probability vector $\vect{p}_k$, $\mathcal{O}(\vect{p}_k)$ can be evaluated by estimating the expected user rates $\bar{R}_i(\vect{p}_k), \forall i \in \mathcal{N}$, over an observation window. Therefore, an initial data set of $H_{t=0}$ samples, $\mathcal{D}_0 =\{ \vect{p}_k,{\mathcal{O}}(\vect{p}_k) \}_{k=1}^{H_0}$, can be generated by independently and uniformly sampling $\vect{p}_k, \forall k \in [H_0]$ from $(0,1]^N$, and estimating the objective function values ${\mathcal{O}}(\vect{p}_k)$ at these samples. Based on $\mathcal{D}_0$, a compatible surrogate model for $\mathcal{O}(\vect{p})$ is learned in a system identification stage. Next, in the system control stage, the learned surrogate model is used to sample new data points $\vect{p}_k, \forall k \in [H_t]$, in a way to strike a trade-off between model exploration and model exploitation. The system identification and system control stages are iterated until the optimal transmission probability vector $\vect{p}^*$ is found, or the total sampling budget $H = \sum_{t=0} H_t$ is exhausted. In the following sub-sections, the system identification and system control stages are discussed in detail, and the overall active-learning-based optimization algorithm is presented.

\subsection{System Identification}
In model-based system identification, $\mathcal{O}(\vect{p})$ is modeled by $\mathcal{O}(\vect{p}) = \hat{\mathcal{O}}_\mu(\vect{p};\vect{\theta}) + \varepsilon(\vect{p};\vect{\theta})$, where $\hat{\mathcal{O}}_\mu(\vect{p};\vect{\theta})$ is a parameterized model which maps the transmission probability vector $\vect{p}$ and the set of model parameters $\vect{\theta}$ to the expected value of $\mathcal{O}(\vect{p})$, and $\varepsilon(\vect{p};\vect{\theta}) \sim \mathcal{N}\big(0,\sigma^2(\vect{p};\vect{\theta})\big)$ is a gaussian random variable with zero mean and variance $\sigma^2(\vect{p};\vect{\theta})$ to account for the heteroscedastic estimation errors in the observations. The optimal vector of model parameters $\vect{\theta}^*$ is estimated based on the current dataset $\mathcal{D}_T = \cup_{t=0}^T~ \mathcal{D}_t$ using Maximum Likelihood estimation,
 \begin{equation} \label{eq:MLest}
 \begin{aligned}
     \vect{\theta}^* &=\underset{\vect{\theta}}{\text{argmax}}~\text{log}\Big(\text{Pr}\big(  {\mathcal{O}}(\vect{p}) \big| \vect{p}  ; \vect{\theta}, \mathcal{D}_T \big) \Big)\\
    & =\underset{\vect{\theta}}{\text{argmax}}~\text{log}\Big(\prod_{k} \text{Pr}\big(  {\mathcal{O}}(\vect{p_k}) \big|  \vect{p}_k  ; \vect{\theta} \big) \Big)\\
      & =\underset{\vect{\theta}}{\text{argmax}}~ \sum_{k} \text{log}\Big(\text{Pr}\big(  {\mathcal{O}}(\vect{p_k}) \big|  \vect{p}_k  ; \vect{\theta} \big) \Big) \\
       & =\underset{\vect{\theta}}{\text{argmin}}~ \frac{1}{2} \sum_{k} \Big[\frac{ \mathcal{O}(\vect{p}_k) - \hat{\mathcal{O}}_\mu(\vect{p}_k;\vect{\theta}) }{\sigma(\vect{p}_k;\vect{\theta})}\Big]^2 + \\
       &~~~~~~~~~~~~~~~~~~~~~~~~~\frac{1}{2} \sum_{k}  \text{log}\big(\sigma^2(\vect{p}_k;\vect{\theta}) \big),
\end{aligned}
 \end{equation}  
 i.e., the optimal surrogate model for $\mathcal{O}(\vect{p})$ can be learned through regression by minimizing the negative of the log-likelihood function. Notice that minimizing the negative log-likelihood criterion jointly minimizes the sum of squared prediction errors between the sample value $\mathcal{O}(\vect{p}_k)$ and the mean predicted value $\hat{\mathcal{O}}_\mu(\vect{p}_k;\vect{\theta})$ of the model, as well the variance of the predicted value $\sigma^2(\vect{p}_k;\vect{\theta})$ in the current dataset $\mathcal{D}_T$ \cite{ljung1999system}. The results of theorem 1, theorem 2, and our conjecture that $\mathcal{O}(\vect{p})$ is a strictly concave function on a convex set $S^N$ provide a key insight for efficient system identification as we can restrict the space of compatible models for $\mathcal{O}(\vect{p})$ to the class of concave functions. By learning an accurate concave model $\hat{\mathcal{O}}_\mu(\vect{p};\vect{\theta}^*)$, the globally optimal transmission probability vector $\vect{p}^*$ can be found using common convex optimization techniques, i.e., $\vect{p}^* = \underset{\vect{p}}{\text{argmax}}~\hat{\mathcal{O}}_\mu(\vect{p};\vect{\theta}^*)$. 
 
\begin{figure}
    \centering
    \includegraphics[width=1\linewidth]{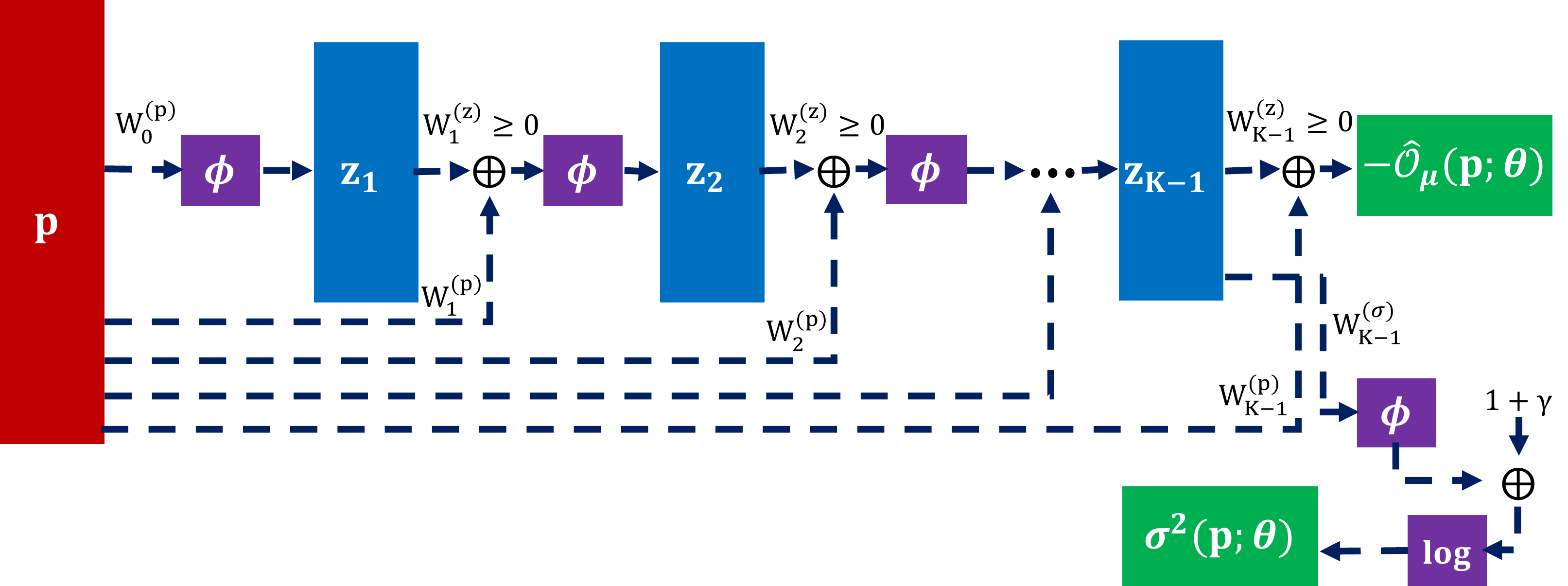}
    \caption{Input Convex (Concave) Feed-Forward Neural Network } \label{ICNN}
    \vspace{-5mm}
 \end{figure}

 Deep neural networks (DNNs) are one of the most commonly adopted models because they are universal function approximators in the sense that they can approximate any continuous function of $N$ variables on a compact domain arbitrarily well \cite{poggio2019theoretical,hornik_approximation}. In DNNs, model parameters $\vect{\theta}$ represent the set of neuron weights and biases. While DNNs have been used for system identification to model high-dimensional systems with nonlinear maps \cite{narendra_approximation, jag_discrete}, their adoption can be challenging for optimal system control because DNNs are typically non-convex with respect to their inputs. To overcome this challenge, we construct DNNs which are convex with respect to their inputs, referred to as Input Convex Neural Networks (ICNNs), by leveraging two facts about the composition  of convex functions \cite{amos2017input,chen2018optimal},
(i) a non-negative weighted sum of convex functions is itself convex, and (ii) composition of a convex function $g(x)$ and convex non-decreasing function $h(x)$ produces a convex function $h(g(x))$. Thus, by restricting the weights in the DNN to be non-negative and by choosing a convex non-decreasing activation function, an ICNN can be designed. In this work, the exponential linear activation function is adopted,
\begin{equation}
\phi(x) =
\begin{cases}
x, ~~~~~~~~~~x \geq 0, \\
\gamma(e^x - 1), x \leq 0,
\end{cases}
\end{equation}
where $\gamma$ is a hyper-parameter that controls the value to which the activation function saturates for negative inputs. Figure \ref{ICNN} illustrates a general architecture of a feed-forward ICNN. Specifically, this model defines a deep neural network over the input $\vect{p}$ using the architecture for layers $k=0,1, \cdots, K-1$,
\begin{equation}
\begin{aligned}
    &\vect{z}_{k+1} = \phi\Big(\vect{W}_k^{(\vect{z})}\vect{z}_k + \vect{W}_k^{(\vect{p})} \vect{p} + \vect{b}_k \Big), \\
    &\hat{\mathcal{O}}_\mu(\vect{p}; \vect{\theta}) = -\Big( \vect{W}_{K-1}^{(\vect{z})}\vect{z}_{K-1} + \vect{W}_{K-1}^{(\vect{p})} \vect{p} + \vect{b}_{K-1} \Big), \\
    &\sigma^2(\vect{p};\vect{\theta}) = \text{log}\Big(1+\gamma+\phi \Big(\vect{W}^{(\sigma)}\vect{z}_{K-1} + \vect{b}_{\sigma}\Big) \Big),
\end{aligned}
\end{equation}
where $\vect{W}_k^{(\vect{z})} \geq 0$ is a matrix of non-negative neuron weights for the $k$-th layer (with $\vect{z}_0,\vect{W}_0^{(\vect{z})}=\vect{0}$), $\vect{W}_k^{(\vect{p})}$ is a matrix of neuron weights for the $k$-the direct pass-through layer $k$,  $\vect{b}_k$ is a column vector of neuron biases, and $\vect{\theta} = \big\{\vect{W}_{1:K-1}^{(\vect{z})}, \vect{W}_{0:K-1}^{(\vect{p})}, \vect{b}_{0:K-1},\vect{W}^{(\sigma)}, \vect{b}_\sigma  \big\}$ is the set of ICNN model parameters. Notice that the ICNN network outputs two values in the final layer, corresponding to the predicted mean   $\hat{\mathcal{O}}_\mu(\vect{p}; \vect{\theta})$ and variance $\sigma^2(\vect{p};\vect{\theta}) \geq 0$. Compared with a standard feed-forward DNN, an ICNN has the notable addition of the direct pass-through layers $\vect{W}_{0:K-1}^{(\vect{p})}$ which directly connect the input $\vect{p}$ to the hidden units in deep layers. These layers are necessary in ICNNs because the non-negativity constraints on $\vect{W}_{1:K-1}^{(\vect{z})}$  restricts previous hidden units from being mapped to subsequent hidden units with the identity mapping, and so explicitly including these direct pass-through layers enables the model to represent identity mappings. Notice that $\vect{W}_{0:K-1}^{(\vect{p})}$ can take any value because $\vect{W}_{k}^{(\vect{p})} \vect{p}$ are just linear functions with respect to $\vect{p}$ and hence convex. The output neuron corresponding to the predicted mean $\hat{\mathcal{O}}_\mu(\vect{p}; \vect{\theta})$ is multiplied by a negative sign to make the ICNN-based model a concave function of $\vect{p}$. Because the variance $\sigma^2(\vect{p};\vect{\theta})$ can be a non-convex function of $\vect{p}$, an additional layer is applied to $\vect{z}_{K-1}$ in which $\vect{W}^{(\sigma)}$ can take any value, such that $\sigma^2(\vect{p};\vect{\theta})$ can model any non-convex function. 

By learning $\sigma^2(\vect{p};\vect{\theta})$, the ICNN-based model captures aleatoric uncertainty, which is the noise inherent in the observations with respect to the inputs $\vect{p}$ \cite{kendall2017uncertainties}. To capture epistemic uncertainty, which is uncertainty in the model
parameters due to scarcity of the data in some regions, we train an ensemble of ICNN models, where each ICNN model is trained on the current data set $\mathcal{D}_T$ with random initialization of the ICNN parameters \cite{lakshminarayanan2017simple}. The ensemble is treated as a uniformly-weighted Gaussian mixture model with mean and variance of the mixture given by, 
\begin{equation}
\resizebox{1\hsize}{!}{$
    \begin{aligned}
    &\hat{\mathcal{O}}_{\mu_*}(\vect{p}) = \frac{1}{E} \sum_{m=1}^E \hat{\mathcal{O}}_{\mu_m}(\vect{p};\vect{\theta}_m), \\
    &\sigma^2_*(\vect{p}) = \frac{1}{E}\sum_{m=1}^E \big(\sigma^2_m(\vect{p};\vect{\theta}_m) + (\hat{\mathcal{O}}_{\mu_m}(\vect{p};\vect{\theta}_m))^2\big) - (\hat{\mathcal{O}}_{\mu_*}(\vect{p}))^2, 
    \end{aligned} $}
\end{equation}
where $E$ is the number of models in the ensemble, and {$\hat{\mathcal{O}}_{\mu_m}(\vect{p};\vect{\theta}_m)$ and $\sigma^2_m(\vect{p};\vect{\theta}_m)$ are the predicted mean and variance of the individual models in the ensemble, respectively.} It is worth mentioning that epistemic uncertainty can be reduced by collecting more data points $\{\vect{p}_k,{\mathcal{O}}(\vect{p}_k) \}_{k}$ in regions where $\sigma^2_*(\vect{p})$ is high.  



\subsection{System Control}
By capturing both aleatoric and epistemic uncertainties, we can design an asymptotically competitive model-based learning algorithm in which the surrogate model is sequentially improved and optimized by sampling new data points in a way to strike a balance between model exploitation and model exploration. Specifically, in model exploitation, a new data point $\{p_k,\mathcal{O}(\vect{p_k})\}$,
where $\vect{p}_k = \underset{\vect{p}}{\text{argmax}}~\hat{\mathcal{O}}_{\mu_*}(\vect{p})$, is sampled based on the current surrogate model in an attempt to maximize the true $\mathcal{O}(\vect{p})$. Notice that $\vect{p}_k$ can be found using common convex optimization techniques since $\hat{\mathcal{O}}_{\mu_*}(\vect{p})$ is a convex function of $\vect{p}$ by design. In model exploration, a new set of new data points $\mathcal{D}_t=\{p_k,\mathcal{O}(\vect{p_k})\}_{k=1}^{H_t}$ are sampled to reduce the predictive uncertainty of the learned surrogate model by $\vect{p}_k = \underset{\vect{p}}{\text{argmax}}~\sigma^2_*(\vect{p})$. Because $\sigma^2_*(\vect{p})$ is a non-convex function, a local derivative-free optimizer such as Nelder-Mead \cite{nelder1965simplex} can be started from a set of $H_t$ points which are independently and uniformly sampled from $\vect{p}_k \in (0,1]^N$, in order to find $H_t$ local maxima of $\sigma^2_*(\vect{p})$. Next, $\mathcal{O}(\vect{p_k})$ is estimated at the set of local maxima of $\sigma^2_*(\vect{p})$ to generate the new data set $\mathcal{D}_t$. By augmenting $\mathcal{D}_t$ to the existing data set and re-training the ICNN-based models, the predictive uncertainty of the surrogate model is improved. 

Model exploitation and model exploration can be combined in one-step by sampling points $\vect{p}_k$ which maximize an upper confidence bound (UCB), $\vect{p}_k = \underset{\vect{p}}{\text{argmax}}~\hat{\mathcal{O}}_{\mu_*}(\vect{p})+\beta \sigma_*(\vect{p})$, where $\beta \geq 0$ is a non-negative parameter to make a trade-off between model exploration and model exploitation. Therefore, by successively sampling new data points to maximize the UCB of the surrogate model (system control), and updating the surrogate model based on the new samples (system identification), the optimal transmission probability vector $\vect{p}^*$ which maximizes \eqref{eq:system} can be found asymptotically. In practice, system control and system identification stages are iterated until the total sampling budget $H = \sum_{t=0} H_t$ is exhausted, and $\vect{p}_k = \underset{k\in [H]}{\text{argmax}}~\mathcal{O}(\vect{p}_k)$ is considered to be a (sub)optimal solution. 

\subsection{The centralized learning algorithm}
To execute the proposed centralized learning-based optimization algorithm, a back-end central controller successively constructs the data set $\mathcal{D}_T$ and iterates through the system control and system identification stages as shown in Algorithm 1, until the optimal transmission probability vector $\vect{p}^*$ is found or the sampling budget is exhausted. 
By exploiting the structure of the optimization problem and trading-off model exploration and model exploitation, Algorithm 1 offers an asymptotically efficient data-driven solution for optimizing the mathematically intractable objective of \eqref{eq:system}. Yet it may not scale well with the number of IoT devices due to the curse of dimensionality in learning high-dimensional models based on a finite number of samples.
To overcome these challenges, we further study the design of a scalable distributed learning-based optimization algorithm in the following section.

\begin{algorithm} \label{Alg1}
 \SetKwInOut{Input}{Input}
 \SetKwInOut{Output}{Output}
    \Input{
    Initial data set $\mathcal{D}_0 =\{ \vect{p}_k,{\mathcal{O}}(\vect{p}_k) \}_{k=1}^{H_0}$\\
    Sampling budget vector $(H_1,\cdots,H_T)$ \\
    Exploration parameter schedule $(\beta_1,\cdots,\beta_T)$
    }
    \Output{(Sub)optimal transmission probability vector $\vect{p}^*$ }
Set $\mathcal{D} = \mathcal{D}_0$\\
\For{$t=1,2 \cdots, T$}
{
Train an ensemble of ICNN-based models based on $\mathcal{D}$ using \eqref{eq:MLest}\\
Find $H_t$ transmission probability vectors $\vect{p}_k = \underset{\vect{p}}{\text{argmax}}~\hat{\mathcal{O}}_{\mu_*}(\vect{p})+\beta_t \sigma_*(\vect{p}), k\in[H_t]$, by starting a local optimizer from $H_t$ random points\\
Estimate \eqref{eq:system} at the new set of data points to generate $\mathcal{D}_t=\{p_k,\mathcal{O}(\vect{p_k})\}_{k=1}^{H_t}$\\
Augment new data set to training data set $\mathcal{D}= \{\mathcal{D},\mathcal{D}_{t}\}$\\
}
$\vect{p}^* = \underset{k\in \mathcal{D}}{\text{argmax}}~ \mathcal{O}(\vect{p}_k)$
    \caption{Centralized Learning-Based Network Optimization}
\end{algorithm}

\section{Distributed Learning-Based Optimization Algorithm}

To design a scalable distributed learning-based optimization algorithm in which each individual IoT device adapts it transmission probability $p_i, \forall i \in \mathcal{N}$, we first reformulate the problem of optimizing \eqref{eq:system} as an $N$-player strategic  game $\mathcal{G}=\langle \mathcal{N},\mathcal{A},\mathcal{U} \rangle$ which consists of,
\begin{enumerate}
    \item \textbf{Players}: the set of players is the set of IoT devices $\mathcal{N}$. Each IoT device in the game $\mathcal{G}$ is a rational decision maker who makes decisions autonomously. 
    \item \textbf{Action space}: the action space for each player $i$ is the compact convex set of transmission probabilities $\mathcal{A}_i = \{p_i |~ p_i \in [\varepsilon_p,1]\}$, where $\varepsilon_p$ is 
    a small positive infinitesimal quantity to make the action space compact. The strategy profile $\mathcal{A}= \mathcal{A}_1 \times \cdots \mathcal{A_N}$
    is a set of strategies for all players which fully specifies all actions in the game. 
    \item \textbf{Utility functions}: a utility function of player $i$ is a function of the strategy profile of all players $\mathcal{U}_i : \mathcal{A} \rightarrow \mathbb{R}$. Every player $i$ selects an action $p_i$ considering the possible actions of all other players $\vect{p}_{-i} = (p_1,\cdots, p_{i-1}, p_{i+1}, \cdots, p_{N})$ to maximize its own utility $\mathcal{U}_i(p_i,\vect{p}_{-i})$. Let $\mathcal{U}$ be the utility profile of all players, $\mathcal{U}=\big(\mathcal{U}_1,\cdots, \mathcal{U}_N \big)$. In $\mathcal{G}$, all players have the same utility function  $\mathcal{U}_i(\vect{p}) = \mathcal{O}(\vect{p}), \forall i \in \mathcal{N}$, as given by \eqref{eq:system}.
\end{enumerate}

A Nash equilibrium solution is the canonical solution for strategic games, which ensures a stable state of the game. Specifically, in a strategic game  $\langle \mathcal{N},\mathcal{A},\mathcal{U} \rangle$ of $N$-players, the set of strategy profiles $\vect{p}^* \in \mathcal{A}$ constitute a set of Nash equilibria if $\forall i, p_i \in \mathcal{A}_i$, $\mathcal{U}_i(p_i^*, \vect{p}_{-i}^*) \geq \mathcal{U}_i(p_i, \vect{p}_{-i}^*)$,
i.e., a Nash equilibrium solution is an action profile $\vect{p}^* \in \mathcal{A}$ of all players with the property that no player $i$ can improve its utility $\mathcal{U}_i(\vect{p}^*)$ by unilaterally choosing an action $p_i$ different from $p_i^*$ given that every other player $j$ adheres to $p_j^* \in \mathcal{A}_j, \forall j \in \mathcal{N}\setminus\{ i\}$. 

\begin{definition}
A game in which the strategy sets are compact convex sets and the utility functions $\mathcal{U}_i: \mathcal{A}\rightarrow \mathbb{R}, \forall i \in \mathcal{N}$ are continuously differentiable is a continuous potential game if there exists continuously differentiable function $\mathcal{P}: \mathcal{A}\rightarrow \mathbb{R}$ such that \cite{monderer1996potential}, 

\begin{equation}
    \frac{\partial \mathcal{P}(p_i, \vect{p}_{-i})}{\partial p_i} =  \frac{\partial \mathcal{U}_i(p_i,\vect{p}_{-i})}{\partial p_i}, \forall i \in \mathcal{N}
\end{equation}

\end{definition}

\begin{proposition}
The strategic game $\mathcal{G}$ is a continuous potential game.
\end{proposition} 

\begin{proof*}
Notice that the strategy profile of all players in $\mathcal{G}$ is the compact convex set $\mathcal{A}$. 
Because all players have the same continuous utility function $\mathcal{U}_i(p_i, \vect{p}_{-i}) = \mathcal{O}(p_i,\vect{p}_{-i}), \forall i \in \mathcal{N}$, $\mathcal{P}(\vect{p})=\mathcal{O}(\vect{p})$ is a potential function for the game $\mathcal{G}$. Hence, $\mathcal{G}$ is a continuous potential game.
 $\blacksquare$
\end{proof*}

Continuous potential games with compact strategy sets have several useful properties. First, they possess at least one pure strategy Nash equilibrium \cite{monderer1996potential}. Second, every Nash equilibrium is a stationary point of $\mathcal{P}$, which includes the set of local maxima and saddle points of $\mathcal{P}$. If $\mathcal{P}$ is concave and bounded, then every pure Nash equilibrium is a maximum point of $\mathcal{P}$ which coincides with $\underset{\vect{p}}{\text{argmax}}~\mathcal{P}(\vect{p})$ \cite{neyman1997correlated}. Moreover, the pure Nash equilibrium is unique if $\mathcal{P}$ is strictly concave and bounded. Based on these properties and the results of theorems 1 and 2, we can conclude that $\mathcal{P}(\vect{p}) = \mathcal{O}(\vect{p})$ has a unique pure Nash equilibria which is the global maximizer of $\mathcal{P}(\vect{p})$ for the case of $N=2,3$ users, respectively. For a general case of $N>3$, we can conclude that $\mathcal{P}(\vect{p})$ has at least one pure Nash equilibrium. The existence of pure Nash equilibria in the strategic game $\mathcal{G}$ is important because it allows for the possibility to design a distributed algorithm which converges to a stable state when IoT devices unilaterally adapt their transmission probabilities to maximize their own utilities.

Because a pure Nash equilibrium $\vect{p}^* \in \mathcal{A}$ is by definition a fixed 
point of the joint best-response (BR) mapping in which each player plays its best response $\mathcal{B}_i(\vect{p}_{-i}^*)$ to the actions of other players $\vect{p}_{-i}^*$,
\begin{equation}
    p_i^* \in \mathcal{B}_i(\vect{p}_{-i}^*) = \underset{p_i\in[\varepsilon_p,1]}{\text{argmax}}~\mathcal{P}(p_i, \vect{p}_{-i}^*),~~\forall i \in \mathcal{N} ,
\end{equation}
it is natural to design algorithms based on BR mappings so that players adaptively learn to play a pure Nash equilibrium strategy over time. Notice that $\mathcal{B}_i(\vect{p}_{-i})$ is a set-valued function because there may be many local maxima. While BR dynamics offer a simple approach to find a pure Nash equilibrium, in practice however, BR dynamics may cycle and never terminate \cite{voorneveld1999potential,wright2015coordinate}. 
Furthermore, BR dynamics require that every player either have an analytic form for $\mathcal{B}_i(\vect{p}_{-i})$ or be able to optimize $\mathcal{P}(p_i, \vect{p}_{-i})$ given the current strategy profile of all other players $\vect{p}_{-i}$. In the case of $\mathcal{G}$, neither $\mathcal{B}_i(\vect{p}_{-i})$ nor $\mathcal{P}(p_i, \vect{p}_{-i})$ is available 
in closed-form, and collecting the strategy profile of all other players $\vect{p}_{-i}$ will incur significant communication overhead for the algorithm. 

To tackle these challenges, we study the design of a distributed learning-based algorithm in which each player plays a best response strategy to an observable aggregate quantity that is a function of other players' strategies $\mathcal{J}_i(\vect{p}_{-i})$, therefore reducing the communication overhead. While economists have studied the design of action aggregators for simple utility functions and analyzed the convergence of BR learning to Nash equilibria \cite{dubey2006strategic,jensen2010aggregative,acemoglu2013aggregate}, the design of an action aggregator remains a game-specific challenge in general, in the sense that an action aggregator should fully capture the game structure and enable a player to play a convergent BR strategy. In the following sub-section, we design a learning-based BR algorithm with action aggregation for $\mathcal{G}$, and prove its convergence to a Nash equilibrium. 

\subsection{Best Response Dynamics with Action Aggregation}

To design a suitable action aggregator $\mathcal{J}_i(\vect{p}_{-i})$ for $\mathcal{G}$, we first analyze the structure of the potential function $\mathcal{P}$ of the game. Recall that $\bar{R}_i(\vect{p})=\bar{R}_i(p_i, \vect{p}_{-i})$ is the expected rate of user $i$ given its transmission probability $p_i$ and the transmission probability vector of all other players $\vect{p}_{-i}$. Note that $\bar{R}_i(p_i, \vect{p}_{-i})=p_iV_i(\vect{p}_{-i})$, where $V_i(\vect{p}_{-i})$ is the conditional expected rate of user $i$ given that it transmits. To expose the impacts of user $j$ on the achievable expected rate by user $i$, $ \bar{R}_i(p_i,\vect{p}_{-i})$ can also be written as,
\begin{equation} \label{eq:ji_impact}
 \bar{R}_i(\vect{p}) = p_j V_i(\vect{p}_{-j}) + (1-p_j)U_i(  \vect{p}_{-j})
\end{equation}
where $\vect{p}_{-j} = (p_1,\cdots, p_{j-1}, p_{j+1}, \cdots, p_{N})$ is the transmission probability vector of all users except of user $j$'s, $V_i(\vect{p}_{-j})$ is the conditional expected rate of user $i$ given that user $j$ transmits, and $U_i(\vect{p}_{-j})$ is the conditional expected rate of user $i$ given that user $j$ does not transmit. Notice that $U_i(\vect{p}_{-j}) > V_i(\vect{p}_{-j})$ because transmissions of user $j$ cause interference to user $i$'s transmissions and therefore degrade user $i$'s expected rate. 

W.L.O.G., set $i=1$ and consider the potential function $\mathcal{P}$ of the game $\mathcal{G}$ from the perspective of user $1$ given a fixed strategy profile of all other users $\vect{p}_{-1}$. By substituting \eqref{eq:ji_impact} with $i=k,\forall k\neq 1$, in $\eqref{eq:system}$, 
\begin{equation} \label{eq:p_from_i}
\resizebox{1\hsize}{!}{$
\begin{aligned}
&\mathcal{P}(p_1,\vect{p}_{-1}) = \frac{1}{N}\text{log}(\bar{R}_1(p_1, \vect{p}_{-1})) + \frac{1}{N}\sum_{k\neq 1} \text{log}(\bar{R}_k(p_1, \vect{p}_{-1})) \\
&=\frac{1}{N} \text{log}(p_1 V_1(\vect{p}_{-1})) + \frac{1}{N} \sum_{k\neq 1} \text{log}(p_1 V_k(\vect{p}_{-1}) + (1-p_1)U_k(  \vect{p}_{-1})).
\end{aligned}$}
\end{equation}
To design a BR strategy in which user $1$ adapts its transmission probability based on the action aggregator $\mathcal{J}_1(\vect{p}_{-1})$, the second summand in \eqref{eq:p_from_i} is approximated by the $K$-th order Taylor series expansion of the log function,
\begin{equation} \label{eq:approx}
    \resizebox{1\hsize}{!}{$
    \begin{aligned}
& \sum_{k\neq 1} \text{log}(p_1 V_k(\vect{p}_{-1}) + (1-p_1)U_k(  \vect{p}_{-1}))  \\
&=  \sum_{k\neq 1} \text{log}\Big(U_k(  \vect{p}_{-1})\frac{U_k(  \vect{p}_{-1})+p_1( V_k(\vect{p}_{-1}) -U_k(  \vect{p}_{-1}))}{U_k(  \vect{p}_{-1})} \Big) \\
&=  \sum_{k\neq 1} \text{log}\big(U_k(  \vect{p}_{-1}) \big) + \text{log}\Big(1+ p_1\frac{ V_k(\vect{p}_{-1}) -U_k(  \vect{p}_{-1})}{U_k(  \vect{p}_{-1})} \Big) \\
&\approx \sum_{k\neq 1} \Big[ \text{log}\big(U_k(  \vect{p}_{-1}) \big) -  \sum_{n=1}^K \frac{p_1^n \big(Q_k(\vect{p}_{-1})\big)^n}{n} \Big]
    \end{aligned} $}
\end{equation}
where $ Q_k(\vect{p}_{-1})= \frac{U_k(\vect{p}_{-1}) - V_k(\vect{p}_{-1})}{U_k(\vect{p}_{-1})} > 0$. Based on \eqref{eq:p_from_i} and \eqref{eq:approx}, the first and second partial derivatives of $\mathcal{P}(p_1,\vect{p}_{-1})$ with respect to $p_1$ for a given $\vect{p}_{-1}$ are 
\begin{equation}
\resizebox{1\hsize}{!}{$
\begin{aligned}
  \frac{\partial \mathcal{P}(p_1,\vect{p}_{-1})}{\partial p_1}& \approx \frac{1}{N}\Bigg[ \frac{1}{p_1} - \sum_{k\neq 1} \Big[ \sum_{n=1}^K p_1^{n-1}{\big(Q_k(\vect{p}_{-1})\big)^n} \Big] \Bigg]
\end{aligned}$}
\end{equation}
and
\begin{equation}
\resizebox{1\hsize}{!}{$
\begin{aligned}
  \frac{\partial \mathcal{P}^2(p_1,\vect{p}_{-1})}{\partial p_1^2}& \approx \frac{1}{N}\Bigg[ \frac{-1}{p_1^2} - \sum_{k\neq 1}  \Big[ \sum_{n=2}^K (n-1)p_1^{n-2}{\big(Q_k(\vect{p}_{-1})\big)^n} \Big] \Bigg],
\end{aligned} $}
\end{equation}
respectively. Because $\frac{\partial \mathcal{P}^2(p_1,\vect{p}_{-1})}{\partial p_1^2} <0, \forall p_1 \in [\varepsilon_p,1]$,  $\mathcal{P}(p_1,\vect{p}_{-1})$ is a strictly concave function of $p_1$, and there exists at most one strict local maximum $p_1^* \in [\varepsilon_p,1]$, which is also the unique strict global maximum of $\mathcal{P}(p_1,\vect{p}_{-1})$ with respect to $p_1$. If $p_1^*$ exists, it satisfies the first-order optimality condition $\frac{\partial \mathcal{P}(p_1,\vect{p}_{-1})}{\partial p_1}=0$ given by the polynomial equation, 
\begin{equation} \label{eq:root}
\begin{aligned} 
    p_1 &= \frac{1}{\sum_{k\neq 1} \Big[ \sum_{n=1}^K p_1^{n-1}{\big(Q_k(\vect{p}_{-1})\big)^n} \Big]}\\
    & = \frac{1}{ \langle \mathcal{J}_1(\vect{p}_{-1}), (1, p_1,\cdots, p_1^{K-1} ) \rangle }\\
    &= \mathcal{F}_1(p_1)  
\end{aligned}
\end{equation}
where the action aggregator is the vector 
$\mathcal{J}_1(\vect{p}_{-1})$,
\begin{equation}
\begin{aligned}
&\mathcal{J}_1(\vect{p}_{-1}) = \\
&\Big(\sum_{k\neq 1}Q_k(\vect{p}_{-1}), 
\sum_{k\neq 1} \big(Q_k(\vect{p}_{-1}) \big)^2,\cdots, 
\sum_{k\neq 1} \big(Q_k(\vect{p}_{-1}) \big)^K\Big)    
\end{aligned}
\end{equation}
and $\langle \mathcal{J}_1(\vect{p}_{-1}), (1, p_1,\cdots, p_1^{K-1}) \rangle$ denotes the inner product of $\mathcal{J}_1(\vect{p}_{-1})$ with the vector $(1, p_1 ,\cdots, p_1^{K-1})$. The dependence of $\mathcal{F}_1(p_1)$ on a fixed $\vect{p}_{-1} \in \mathcal{A}_{-1}$ has been suppressed for notational convenience. Notice that 
$\sum_{k\neq 1} \Big[ \sum_{n=1}^K p_1^{n-1}{\big(Q_k(\vect{p}_{-1})\big)^n} \Big]  < K(N-1) $, hence $p_1=\mathcal{F}_1(p_1) > \frac{1}{K(N-1)} \geq \varepsilon_p$. The BR strategy of user $1$ to the actions of other players $\vect{p}_{-1}$ is therefore the unique root $p_1^* \in [\varepsilon_p,1]$ of \eqref{eq:root}, or otherwise $p_1^*=1$ in the case when user 1's transmissions are outside the interference range of other users in the network,  
\begin{equation} \label{p:fp}
\begin{aligned}
\Tilde{\mathcal{B}}_1(\vect{p}_{-1}) =\text{min}\big\{\mathcal{F}_1(p_1),1\big\} 
\end{aligned}
\end{equation}

To determine the single-valued BR strategy $\Tilde{\mathcal{B}}_1(\vect{p}_{-1})$ in practice, $\mathcal{J}_1(\vect{p}_{-1})$ should be first estimated over an observation window where $\vect{p}_{-1}$ is held fixed. This can be done in two steps. In the first step, user 1 transmits  with probability $p_1$ over the first half of an observation window and the expected user rates $\bar{R}_k(p_1,\vect{p}_{-1}), \forall k \in \mathcal{N}\setminus \{1\}$ are estimated at their respective BSs. In the second step, user 1 abstains from transmission (i.e. $p_1=0$) in the second half of the observation window, and a new set of expected user rates $\bar{R}_k(0,\vect{p}_{-1}), \forall k \in \mathcal{N}\setminus \{1\}$ are estimated. $V_k(\vect{p}_{-1})$ and $U_k(\vect{p}_{-1})$ can be then obtained for every user $k \in \mathcal{N}\setminus \{1\}$ by solving the following system of linear equations, 
\begin{equation} \label{eq:lsystem}
\begin{aligned}
 &\bar{R}_k(p_1,\vect{p}_{-1}) = p_1 V_k(\vect{p}_{-1}) + (1-p_1)U_k(  \vect{p}_{-1}) \\
 &\bar{R}_k(0,\vect{p}_{-1}) = U_k(  \vect{p}_{-1}), 
\end{aligned}
\end{equation}
and so $Q_k(\vect{p}_{-1}), \forall k \in \mathcal{N}\setminus \{1\}$ can be computed and aggregated to obtain $\mathcal{J}_1(\vect{p}_{-1})$. Next, the roots of $p_1 = \mathcal{F}_1(p_1)$ should be found and $\Tilde{\mathcal{B}}_1(\vect{p}_{-1})$ is either $p_1^* = \mathcal{F}_1(p_1^*) \in [\varepsilon_p,1]$ or $p_1^*=1$. Finding the roots of high order polynomials, however, is challenging in practice. For instance, by the Abel–Ruffini theorem, there is no algebraic solution in radicals to general polynomial equations of degree $K\geq 5$ with arbitrary coefficients. Rather than finding the set of all roots, it suffices to find the root $p_1^* \in [\varepsilon_p,1]$ if it exists. To this end, we propose a method based on fixed point iteration in which user $i=1$ find its optimal transmission probability using the sequence,
\begin{equation} \label{ep:fp}
\begin{aligned}
p_i^{n+1} = \Tilde{\mathcal{B}}_i(p_i^n,\vect{p}_{-i}) = \text{min}\big\{ \mathcal{F}_i(p_i^n),1 \big\}.  
\end{aligned}
\end{equation}

Analyzing the convergence \eqref{ep:fp} requires the degree of the polynomial $K$ to be specified. Notice that the higher the order of Taylor approximation $K$ is, the better the approximation of $\mathcal{P}(p_1, \vect{p}_{-1})$ \eqref{eq:p_from_i}. However, increasing $K$ also increases the communication overhead as the aggregation vector $\mathcal{J}_1(\vect{p}_{-1})$ is of dimension $K$. To balance computational accuracy with communication overhead, $K$ is set to $5$. Before presenting convergence analysis of \eqref{ep:fp}, we first introduce the  definition of a Lipschitz function and prove an intermediate lemma regarding the composition of Lipschitz function with point-wise minimum.

\begin{definition}
Let a function $f: [a,b] \rightarrow \mathbb{R}_+$ be such that for some constant $L$  and $\forall x,y \in [a,b]$, 
$$ |f(x)-f(y)| \leq L|x-y|, $$
then the function $f$ is called a Lipschitz function on $[a,b]$, and the least constant $q= \underset{x\neq y }{\text{sup}} \frac{|f(x)-f(y)|}{|x-y|}$ is called the Lipschitz constant.
\end{definition}

\begin{lemma}
Suppose a function $f(x): x\in [a,b] \rightarrow \mathbb{R}_+$ has a Lipschitz constant $q$, then $g(x) = \text{min} \{f(x), 1\}$ has a Lipschitz constant that is at most $q$.  
\end{lemma}
\begin{proof*}
Notice that $g(x) = \text{min} \{f(x), 1\} = \frac{f(x) + 1 - |f(x)-1|}{2}$. Then, 
\begin{equation}
\begin{aligned}
    |&g(x) - g(y)| \\&= \Big|\frac{f(x) - f(y)+|f(y)-1| - |f(x)-1| }{2} \Big| \\
    &\leq \Big|\frac{ f(x) - f(y)+|f(y)- f(x)| }{2} \Big| \\
    &\leq  |f(x) - f(y)|
\end{aligned}
\end{equation}
where the first inequality is obtained based on the reverse triangle inequality $|x|-|y| \leq |x-y|$, and the second inequality is obtained based on the fact that $x-y \leq |x-y|$. Since $f(x)$ has a Lipschitz constant $q$, we have that, 
\begin{equation}
  |g(x) - g(y)|  \leq  |f(x) - f(y)| \leq q |x-y|,
\end{equation}
i.e., the Lipschitz constant of $g(x)$ is at most $q$. $\blacksquare$
\end{proof*}
\begin{lemma}
Given that $Q_k(\vect{p}_{-1})>0, \forall \vect{p}_{-1} \in \mathcal{A}_{-1}$, $\big[\sum_{k\neq 1} Q_k(\vect{p}_{-1}) \big]^2 > \sum_{k\neq 1} \big(Q_k(\vect{p}_{-1})\big)^2 , \forall N \geq 3$
\end{lemma}
\begin{proof*}
The general case of $N\geq 3$ can be proved by induction. For the base case of $N=3$, $\big[Q_2(\vect{p}_{-1}) +Q_3(\vect{p}_{-1}) \big]^2 = \big(Q_2(\vect{p}_{-1}) \big)^2 + \big(Q_2(\vect{p}_{-1}) \big)^2 + 2Q_2(\vect{p}_{-1})Q_3(\vect{p}_{-1}) > \big(Q_2(\vect{p}_{-1})\big)^2 +\big(Q_3(\vect{p}_{-1})\big)^2$. For the induction step, suppose that $\big[\sum_{k\neq 1} Q_k(\vect{p}_{-1}) \big]^2 > \sum_{k\neq 1} \big(Q_k(\vect{p}_{-1})\big)^2$ is true for $N=N^\prime$, then for  $N=N^\prime+1$, $\big[\sum_{k\neq 1} Q_k(\vect{p}_{-1}) +Q_{N^\prime+1}(\vect{p}_{-1}) \big]^2=\big[\sum_{k\neq 1} Q_k(\vect{p}_{-1})\big]^2 + \big( Q_{N^\prime+1}(\vect{p}_{-1})\big)^2 + 2\big[\sum_{k\neq 1} Q_k(\vect{p}_{-1})\big]Q_{N^\prime+1}(\vect{p}_{-1}) > \sum_{k\neq 1} \big(Q_k(\vect{p}_{-1})\big)^2 + \big( Q_{N^\prime+1}(\vect{p}_{-1})\big)^2$ since $2\big[\sum_{k\neq 1} Q_k(\vect{p}_{-1})\big]Q_{N^\prime+1}(\vect{p}_{-1})>0$. $\blacksquare$

\end{proof*}

The convergence of \eqref{ep:fp} is analyzed in proposition 2 and its proof therein. 

\begin{proposition}
Starting from an arbitrary $p_1^0 \in [\varepsilon_p,1]$, the sequence $\{p_1^n\}$ defined by \eqref{ep:fp}, i.e., $p_i^{n+1} = \Tilde{\mathcal{B}}_i(p_i^n,\vect{p}_{-i}) = \text{min}\big\{ \mathcal{F}_i(p_i^n),1 \big\}$, converges to a unique fixed point $p_1^*$ for $n\geq1$ if the number of users is $N \geq 3$. 
\end{proposition}
\begin{proof*} The map governing the dynamics of \eqref{ep:fp} is $\mathcal{T}: p_1 \rightarrow p_1^\prime$, where $p_1^\prime = \Tilde{\mathcal{B}}_1(p_1, \vect{p}_{-1})$.
Notice that $\mathcal{T}: p_1 \in \mathcal{A}_1 \rightarrow p_1^\prime \in \mathcal{A}_1$, i.e., $p_1, p_1^\prime \in [\varepsilon_p,1]$. Define a distance metric $d(p_1,p_1^{\prime \prime})$ $\forall p_1, p_1^{\prime \prime} \in \mathcal{A}_1$ as the $\text{L}_1$ norm, $d(p_1,p_1^{\prime \prime}) = |p_1-p_1^{\prime \prime}|$. Then, by definition, the map $\mathcal{T} : \mathcal{A}_1 \rightarrow \mathcal{A}_1$ 
on the complete metric space $(\mathcal{A}_1,d)$ is called a contraction mapping on $\mathcal{A}_1$ if there exists $q_1 \in [0, 1)$ such that $d(\mathcal{T}(p_1),\mathcal{T}(p_1^{\prime \prime})) \leq q_1d(p_1,p_1^{\prime \prime}), \forall p_1, p_1^{\prime \prime} \in \mathcal{A}_1$. That is if, 
\begin{equation}
\begin{aligned}
d(\mathcal{T}(p_1),\mathcal{T}(p_1^{\prime \prime})) &=  |\Tilde{\mathcal{B}}_1(p_1, \vect{p}_{-1})-\Tilde{\mathcal{B}}_1(p_1^{\prime \prime}, \vect{p}_{-1})| \\
&\leq q_1 |p_1 - p_1^{\prime \prime}|, \forall \vect{p}_{-1}\in \mathcal{A}_{-1},~~~ q<1.
\end{aligned}
\end{equation}
To prove that such $q_1 \in [0, 1)$ exits, we first show that the Lipschitz constant of $\mathcal{F}_1(p_1)$ is strictly less than $1$. To this end, we investigate the maximum absolute rate of change of $\mathcal{F}_1(p_1)$ with respect to $p_1$ given a fixed strategy profile of all other players $\vect{p}_{-1}$, 
\begin{equation} \label{eq:sup}
   \underset{p_1}{\text{sup}} \left\lvert \frac{\partial \mathcal{F}_1(p_1)}{ \partial p_1}\right\rvert = \underset{p_1}{\text{sup}} \frac{\sum_{k\neq 1} \sum_{n=2}^5 (n-1)p_1^{n-2}{\big(Q_k(\vect{p}_{-1})\big)^n}  }{\Bigg[\sum_{k\neq 1} \Big[ \sum_{n=1}^5 p_1^{n-1}{\big(Q_k(\vect{p}_{-1})\big)^n} \Big] \Bigg]^2}. 
\end{equation}
Notice that, 
\begin{equation}
\resizebox{1\hsize}{!}{$
\begin{aligned}
&\Bigg[\sum_{k\neq 1} \Big[ \sum_{n=1}^5 p_1^{n-1}{\big(Q_k(\vect{p}_{-1})\big)^n} \Big] \Bigg]^2\\
&=\Bigg[\sum_{k\neq 1} Q_k(\vect{p}_{-1}) + p_1 Q_k^2(\vect{p}_{-1}) + p_1^2 Q_k^3(\vect{p}_{-1}) + p_1^3 Q_k^4(\vect{p}_{-1}) + p_1^4Q_k^5(\vect{p}_{-1})\Bigg]^2  \\
&>\Bigg[\sum_{k\neq 1} Q_k(\vect{p}_{-1}) \Bigg]^2 + \Bigg[\sum_{k\neq 1} p_1Q_k^2(\vect{p}_{-1}) + p_1^2Q_k^3(\vect{p}_{-1}) \Bigg]^2 \\
&~~~~~~~~~~~~~~+ 2\Bigg[\sum_{k\neq 1} Q_k(\vect{p}_{-1}) \Bigg]\Bigg[ \sum_{k\neq 1} p_1Q_k^2(\vect{p}_{-1}) +p_1^2Q_k^3(\vect{p}_{-1}) +p_1^3Q_k^4(\vect{p}_{-1}) \Bigg] \\
&>\Bigg[\sum_{k\neq 1} Q_k(\vect{p}_{-1}) \Bigg]^2 + \Bigg[\sum_{k\neq 1} p_1Q_k^2(\vect{p}_{-1}) \Bigg]^2 + 2\Bigg[\sum_{k\neq 1} p_1Q_k^2(\vect{p}_{-1}) \Bigg]\Bigg[\sum_{k\neq 1} p_1^2Q_k^3(\vect{p}_{-1}) \Bigg]\\
&~~~~~~~~~~~~~~+
2\Bigg[\sum_{k\neq 1} p_1Q_k^3(\vect{p}_{-1}) \Bigg]  + 2\Bigg[\sum_{k\neq 1} p_1^2Q_k^4(\vect{p}_{-1}) \Bigg] +
2\Bigg[\sum_{k\neq 1} p_1^3Q_k^5(\vect{p}_{-1}) \Bigg] \\
&>\Bigg[\sum_{k\neq 1} Q_k(\vect{p}_{-1}) \Bigg]^2 + 2\Bigg[\sum_{k\neq 1} p_1Q_k^3(\vect{p}_{-1}) \Bigg] + 3\Bigg[\sum_{k\neq 1} p_1^2Q_k^4(\vect{p}_{-1}) \Bigg] + 
4\Bigg[\sum_{k\neq 1} p_1^3Q_k^5(\vect{p}_{-1}) \Bigg],
\end{aligned}$}
\end{equation}
where the strict inequalities come from the fact that some positive terms of the form $p_1^{l^\prime}Q_k^{l^{\prime \prime}}(\vect{p}_{-1})> 0$ for some integers $l^\prime,l^{\prime \prime} $, along with some positive product terms when expanding the parentheses and summations have been dropped. Based on lemma 2, $\big[\sum_{k\neq 1} Q_k(\vect{p}_{-1}) \big]^2 > \sum_{k\neq 1} \big(Q_k(\vect{p}_{-1})\big)^2 , \forall N \geq 3$. This shows that the denominator of \eqref{eq:sup} is strictly larger than its numerator. Hence, $\underset{p_1}{\text{sup}} \left\lvert \frac{\partial \mathcal{F}_1(p_1)}{ \partial p_1}\right\rvert < 1$ and there exists $q_1 \in [0,1)$ for which $\underset{p_1}{\text{sup}} \left\lvert \frac{\partial \mathcal{F}_1(p_1)}{ \partial p_1}\right\rvert \leq q_1$. 

By the mean value theorem from calculus, there exists some $\xi \in (p_1, p_1^{\prime \prime})$ such that $\forall p_1, p_1^{\prime \prime} \in \mathcal{A}_1$
\begin{equation} \label{eq:mvt}
    \mathcal{F}_1^\prime (\xi) =  \frac{\mathcal{F}_1(p_1)-\mathcal{F}_1(p_1^{\prime \prime})}{p_1 - p_1^{\prime \prime}}. 
\end{equation}
By taking the absolute value of both sides in \eqref{eq:mvt} and noting that $|\mathcal{F}_1^\prime (\xi)| \leq q_1$, 
\begin{equation}
    \left\lvert \mathcal{F}_1^\prime (\xi) \right\rvert =  \left\lvert \frac{\mathcal{F}_1(p_1)-\mathcal{F}_1(p_1^{\prime \prime})}{p_1 - p_1^{\prime \prime}} \right\rvert \leq q_1
\end{equation}
i.e., the Lipschitz constant of $\mathcal{F}_1(p_1)$ is $q_1$, which is strictly less than $1$.  Based on the result of lemma 1,  $\forall p_1,p_1^{\prime \prime} \in \mathcal{A}_1$ and $\forall \vect{p}_{-1} \in \mathcal{A}_{-1}$,
\begin{equation} \label{eq:lcineq1}
\begin{aligned}
&\left\lvert \Tilde{\mathcal{B}}_1(p_1,\vect{p}_{-1}) -\Tilde{\mathcal{B}}_1(p_1^{\prime \prime},\vect{p}_{-1})  \right\rvert \leq q^*
\left\lvert{p_1 - p_1^{\prime \prime}}\right\rvert.
\end{aligned} 
\end{equation}
Convergence of \eqref{ep:fp} to a unique fixed point $p_1^* \in [\varepsilon_p,1]$ can be concluded based on Banach fixed point theorem, which guarantees that the contraction mapping $\mathcal{T}$ admits a unique fixed-point $p_1^* \in \mathcal{A}_1$, and that the sequence $\{p_1^0, p_1^2, \cdots\}$ converges to $p_1^*$. $\blacksquare$
\end{proof*}

The analysis and the result of proposition 2 enable the design of a practical discrete-time BR dynamics algorithm in which users sequentially play a best response strategy to the observable action aggregator $\mathcal{J}_i(\vect{p}_{-i}^t)$ of all other users. As shown in Algorithm 2, in each iteration $t\geq0$, one user $i \in \mathcal{N}$, updates its transmission probability using \eqref{ep:fp} to the unique fixed point $p_i^{t*} = \Tilde{\mathcal{B}}_i(p_i^{t*},\vect{p}_{-i}^t)$.  Notice that the algorithm terminates when the change in the common utility function of users is less than a small positive infinitesimal quantity $\varepsilon_T$. The convergence of Algorithm 2 to a pure Nash equilibrium is formalized in Theorem 3. 

\begin{algorithm} \label{Alg2}
 \SetKwInOut{Input}{Input}
 \SetKwInOut{Output}{Output}
    \Input{Instance of the strategic form game $\mathcal{G}=\langle \mathcal{N},\mathcal{A},\mathcal{U} \rangle$\\
    Initial transmission probability vector $\vect{p}^0 \in \mathcal{A}$}
    \Output{Nash equilibrium $\vect{p}^*$ }
Estimate expected user rates $\bar{R}_i(\vect{p}^{0}), \forall i \in \mathcal{N}$\\
\For{$t=0,1, \cdots$, }
{
Randomly pick a user $i \in \mathcal{N}$\\ 
Set $p_i=0$ and estimate expected user rates $\bar{R}_k(0,\vect{p}^{t}_{-i}), \forall k \in \mathcal{N} \setminus \{i\}$\\
Compute $V_k(\vect{p}_{-i}^t)$ and $U_k(\vect{p}_{-i}^t)$ by solving the linear system \eqref{eq:lsystem} using  $\bar{R}_k(p_i^t,\vect{p}^{t}_{-i})$ and $\bar{R}_k(0,\vect{p}^{t}_{-i})$, $\forall k \in \mathcal{N} \setminus \{i\} $ \\
Compute $Q_k(\vect{p}_{-i}^t) = \frac{U_k(\vect{p}_{-i}^t) - V_k(\vect{p}_{-i}^t)}{U_k(\vect{p}_{-i}^t)}, \forall k \in \mathcal{N} \setminus \{i\} $ \\
Compute the action aggregator $\mathcal{J}_i(\vect{p}_{-i}^t)$ \\
Set $p_i^\prime = p_i^t, \varepsilon_{\text{MAE}}=1$\\
\While{$\varepsilon_{\text{MAE}}\geq \varepsilon_T$}
{
$p_i^{\prime \prime}= \text{min}\Big\{\mathcal{F}_i(p_i^\prime),1\Big\}$\\
$\varepsilon_{\text{MAE}}= |p_i^\prime - p_i^{\prime \prime} |$\\
Set $p_i^{\prime } = p_i^{\prime \prime}$
}
Set $p_i^{t+1} = p_i^\prime$\\
Set $\vect{p}^{t+1}=(p_1^t, \cdots, p_{i-1}^t,p_i^{t+1},p_{i+1}^t,\cdots,p_N^t)$\\
Estimate expected user rates $\bar{R}_i(\vect{p}^{t+1}), \forall i \in \mathcal{N}$\\
\If{ $|\mathcal{O}(\vect{p}^{t})- \mathcal{O}(\vect{p}^{t+1})| \leq \varepsilon_T$ }
{Break}
Set $\vect{p}^{t}= \vect{p}^{t+1}$\\
}
    \caption{BR Dynamics with Action Aggregation}
\end{algorithm}

\begin{theorem}
Algorithm 2 converges to a pure Nash equilibrium $\vect{p}^*$ of the strategic form game $\mathcal{G} = \langle \mathcal{N}, \mathcal{A}, \mathcal{U} \rangle$ if the number of players is $N \geq 3$.
\end{theorem}
\begin{proof*}
In every iteration $t$ of Algorithm 2, one user $i \in \mathcal{N}$ deviates from its current strategy $p_i^t$ to $p_i^{t+1}$ such that $p_i^{t+1} = \underset{p_i}{\text{argmax}}~\mathcal{P}(p_i, \vect{p}_{-i}^t)$, where $p_i^{t+1} \in [\varepsilon_p,1]$ is the unique strict global maximizer of $\mathcal{P}(p_i, \vect{p}_{-i}^t)$ given the current strategy profile of all other users $\vect{p}_{-1}^t \in \mathcal{A}_{-i}$, or $p_i^{t+1}=1$ in the case of $\mathcal{F}_i(p_i) > 1, \forall p_i \in [\varepsilon_p, 1)$. If $p_i^{t+1}$ is the strict global maximizer, $\mathcal{P}(p_i^{t+1}, \vect{p}_{-i}^t) > \mathcal{P}(p_i^t, \vect{p}_{-i}^t)$, $\forall p_i^t \neq p_i^{t+1} \in [\varepsilon_p,1]$. In the case of  $\mathcal{F}_i(p_i) > 1$, $\frac{\partial \mathcal{P}(p_i,\vect{p}_{-i}^t)}{\partial p_i}=\frac{1}{N}\Big[\frac{1}{p_i} - \frac{1}{\mathcal{F}_i(p_i)} \Big] > 0, \forall p_i \in [\varepsilon_p,1)$,  and so $\mathcal{P}(p_i^{t+1}=1, \vect{p}_{-i}^t) > \mathcal{P}(p_i^t, \vect{p}_{-i}^t),\forall p_i^t \in [\varepsilon_p,1) $ because $\mathcal{P}(p_i, \vect{p}_{-i}^{t})$ is a strictly monotonously increasing function on $p_i \in [\varepsilon_p,1)$. Hence, a deviation from $p_i^t$ to $p_i^{t+1}$ where $p_i^t\neq p_i^{t+1}$ necessarily improves the potential function of the game $\mathcal{G}$. Because $\mathcal{P} < \infty, \forall \vect{p} \in \mathcal{A}$, Algorithm 2 must terminate when no user can deviate from its current strategy to further improve $\mathcal{P}$. $\blacksquare$  
\end{proof*}

\section{Performance Evaluation}

We have developed a simulator in Python and set up a multi-cell Wireless IoT network with NOMA as described in section \ref{sec:sysmod}, and implemented the proposed centralized and distributed learning algorithms in TensorFlow. Unless mentioned otherwise, the network consists of $N$ IoT devices and $M$ BSs, {where IoT devices are deployed independently and uniformly randomly within a square deployment area of $[-500,500] \times [-500,500]$m, and BSs are deployed in the cluster centroids as determined by Lloyd's ($K=M$)-means clustering algorithm \cite{lloyd1982least} for each realization of the random IoT device deployment.
A Rayleigh fading channel with distance-dependent free-space path-loss is adopted as in \cite{khairy2020constrained}.}
For the centralized learning-based algorithm, an ensemble of $E=10$ ICNN-based models is used, where each ICNN model is a fully connected multi-layer perceptron network as shown in Fig. 1, with three hidden layers of $128$ neurons to predict $-\hat{\mathcal{O}}_\mu(\vect{p}; \vect{\theta})$. An additional hidden layer of $128$ neurons is also used to predict $\sigma^2(\vect{p};\vect{\theta})$. The main simulation parameters used in the experiments are tabulated in Table \ref{table:simparam}. 

\begin{table}[ht]  
\caption{Simulation Parameters}   
\centering                          
\begin{tabular}{c c | c c }            
\hline\hline                        
Parameter  & Value &  Parameter & Value \\ [0.5ex] 
\hline                              
$P_{TX}$ & $30~dBm$  & Hidden layers & $3$    \\
$\alpha$  & $2$  & Neurons per layer &  $128$ \\
$f_0$ & $900$ MHz  & Number of models $E$ & $10$   \\
$d_0$ & $1m$   &$T$ &  $10$   \\
$h_{i,m}$ & $exp(1)$ & $H_t,  \forall t \in [T]$ &  $100$  \\
$n_0$ & $-100dBm$ &  $\beta_t$ &  $\frac{1}{9}(10-t)$\\
$\mathcal{W}$ & $1Hz$  & Dim$[\vect{W}^{(\sigma)}]$ &  $1 \times 128$ \\
$\text{SNIR}_\text{Th}$ & $-5.1dBm$  &  $\epsilon_T$ & $1\times 10^{-6}$ \\
\hline                              
\end{tabular}          \label{table:simparam}   
\end{table}

\subsection{Comparison of Learning Algorithms}
\begin{figure*}[ht]
\centering
\subfloat[$N=4$]{\includegraphics[width=0.33\linewidth]{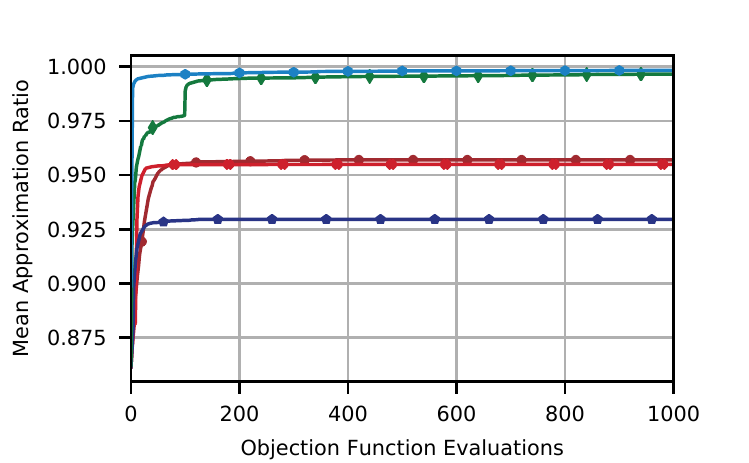}}
\hfill
\centering
\subfloat[$N=8$]{\includegraphics[width=0.33\linewidth]{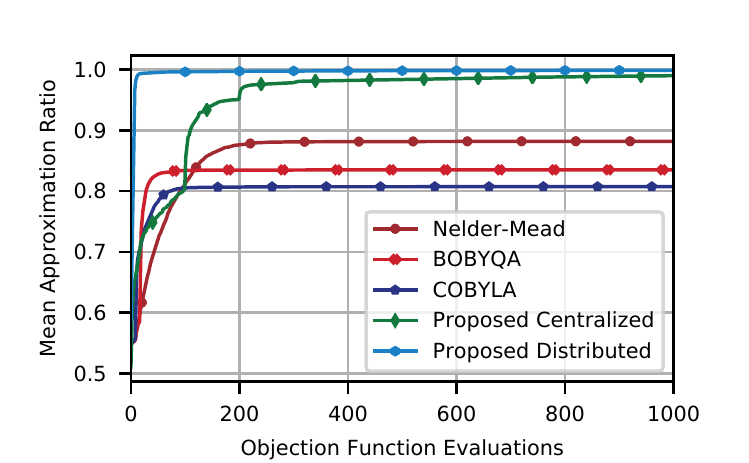}}
\hfill
\centering
\subfloat[$N=16$]{\includegraphics[width=0.33\linewidth]{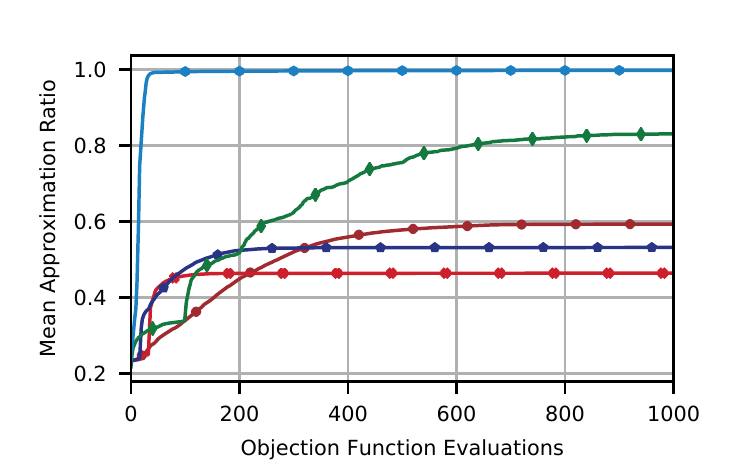}}
\hfill
\caption{Mean approximation ratio comparison of proposed and baseline optimization algorithms on $10$ random deployments of (a) $N=4$, (b) $N=8$, and (c) $N=16$ IoT devices and $M=2$ BSs.} 
\label{fig:optimizerPerformance}
\end{figure*}

\begin{figure}[ht]
\centering
\includegraphics[width=1.0\columnwidth]{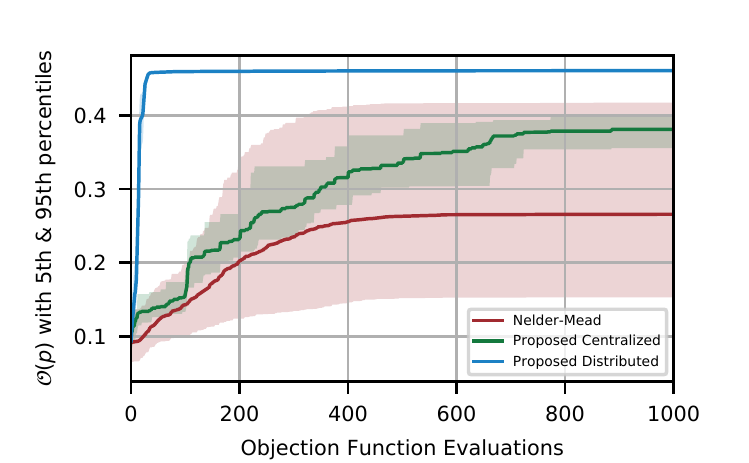}
\caption{Optimization trajectories of different optimization methods on a random topology of $N=16$ IoT devices and $M=2$ BSs.} 
\label{fig:topPerformance}
\end{figure}

\vspace{-1mm}
First, we compare the performance of the proposed centralized and distributed learning algorithms with some common state-of-art derivative-free off-the-shelf optimizers implemented in the NLopt nonlinear optimization package~\cite{nlopt}, namely, BOBYQA~\cite{powell2009bobyqa}, COBYLA~\cite{powell1994direct,powell1998direct}, and Nelder-Mead~\cite{nelder1965simplex}, on $10$ random deployments of a small scale network, i.e., $M=2$ BSs and $N=\{4, 8, 16\}$ IoT devices. The performance of our proposed algorithms is compared with gradient-free methods because analytical gradients of $\eqref{eq:system}$ are not readily available, and estimating the gradient requires at least two evaluations per gradient step for each $p_i,\forall i \in \mathcal{N}$, which makes gradient-based methods noncompetitive in terms of the required number of objective function evaluations. Starting from $10$ random initialization for $\vect{p}^0$, each optimizer is given a budget of $1000$ objective function evaluation to find the optimal transmission probability vector $\vect{p}^*$. In Fig. \ref{fig:optimizerPerformance}, the mean approximation ratio across $100$ runs ($10$ deployments $\times 10$ random initializations) of the proposed and baseline optimization algorithms, is shown as a function of objective function evaluation for the cases of (a) $N=4$, (b) $N=8$, and (c) $N=16$ IoT devices. The approximation ratio is defined as the ratio of the best $\mathcal{O}(\vect{p}^t)$ value attained by an optimizer up to evaluation $t$, to the optimal objective value for each deployment. The optimal objective value for each deployment is the best-found value by any optimization algorithm in any of its $10$ runs. For smaller-scale deployments (e.g. $N=4,8$), the proposed centralized and distributed learning-based optimization algorithms perform comparably in terms of the quality of the solution, although the distributed learning algorithm converges faster. The observation that the distributed learning-based algorithm performs as well as the centralized learning-based algorithm in which a concave surrogate model of $\vect{p}$ is learned, supports our conjecture that \eqref{eq:system} is a concave function of $\vect{p}$ over a convex set $S^N$. The proposed centralized learning-based optimization algorithm however, suffers from the curse of dimensionality as the problem scale increases as evident from Fig. \ref{fig:optimizerPerformance}(c). Both of our proposed machine-learning-based algorithms outperform the commonly used off-the-shelf, derivative-free optimizers such as Nelder-Mead, BOBYQA, COBYLA, which do not exploit any structure present in the optimization problems. In Fig. \ref{fig:topPerformance}, the optimization trajectories of the proposed centralized and distributed learning algorithms, as well as Nelder-Mead (best performing off-the-shelf optimizer) on a random topology of $N=16$ IoT devices and $M=2$ BSs are shown. Each optimizer is randomly initialized $10$ times and the mean value of $\mathcal{O}(\vect{p})$ along with the $5$-th and $95$-th percentiles are plotted. The proposed distributed learning-based optimization algorithm converges to the optimal solution much faster than other algorithms, and exhibits smaller variance in its optimization trajectory. These experiments demonstrate the superiority of our proposed distributed learning-based algorithm compared with other existing algorithms.

\subsection{Network Densification with NOMA}

\begin{figure}[ht]
\centering
\subfloat[$M=2$]{\includegraphics[width=1.0\columnwidth]{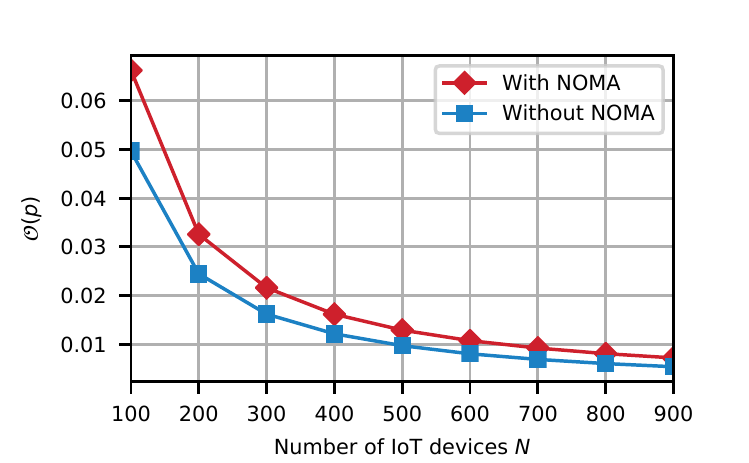}}
\hfill
\centering
\subfloat[$N=200$]{\includegraphics[width=1.0\columnwidth]{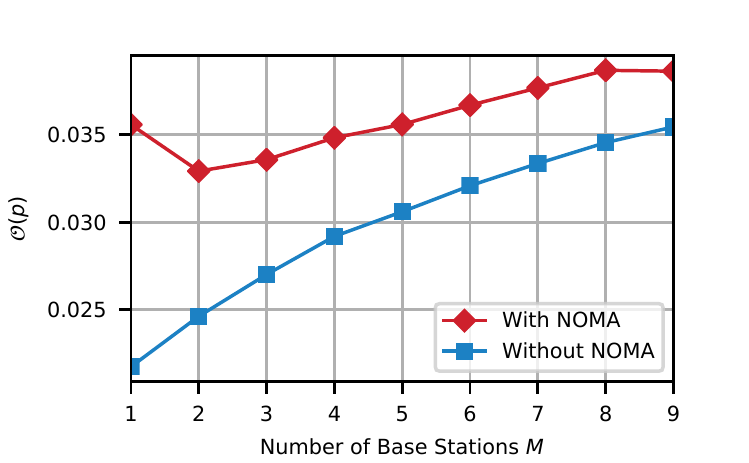}}\\
\caption{Geometric mean of expected user rates with and without NOMA in random deployments of (a) $M=2$ BSs and variable number of IoT devices $N$, and (b) $N=200$ IoT devices and variable number of BSs. 
} 
\label{fig:NomaGain}
\end{figure}

Next, we study the impact of network density with and without NOMA on the geometric mean of expected user rates in Fig. \ref{fig:NomaGain}. Each data point in Figs. \ref{fig:NomaGain}(a) and \ref{fig:NomaGain}(b) represent the mean of the optimal objective function $\mathcal{O}({\vect{p}^*})$ across $10$ random deployments, where the optimal transmission probability vector $\vect{p}^*$ is learned by the distributed learning based algorithm. In Fig. \ref{fig:NomaGain}(a), the number of IoT devices $N$ is varied in $\{100,200,\cdots, 900\}$ while $M$ is fixed to $2$ BSs. The geometric mean of the expected user rates decreases as more users join the network and time-share the wireless broadcast channel. It can also be observed that when the BSs support SIC-based decoding, the expected user rate is increased 
by about $33\%$ compared to the case without NOMA . In Fig. \ref{fig:NomaGain}(b), the number of IoT devices is fixed to $N=200$ while the number of BSs varies in $\{1,2,\cdots, 9\}$. Generally, increasing the number of BSs improves the geometric mean of expected user rates because the impacts of distance-dependent path loss become less severe with more deployed BSs. Notably however, the geometric mean of expected user rates is higher when there is only one NOMA-capable BS, compared with the cases when there are $M=\{2,3,4\}$ NOMA-capable BSs. This is because in a multi-cell network, it is less likely that both of the transmissions with the highest and second highest received SNIRs are from users associated with the same BS compared with single-cell networks. This can also be deduced by evaluating the NOMA gain as a function of $M$. When there is one NOMA-capable BS, NOMA gain is about $63\%$, compared with only $33\%, 16.2\%,8.9\%$ when there are $2,5, 9$ NOMA-capable BSs, respectively. 

\subsection{Heterogeneous Access}

\begin{figure*}[ht]
\centering
\subfloat[$\vect{p}^*$ in an $M=1$ network without NOMA \newline $~~~~~~\text{JF}\big(\bar{R}_1(\vect{p}^*), \cdots,\bar{R}_{144}(\vect{p}^*) \big)=0.9778$]{\includegraphics[width=0.33\linewidth,valign=c]{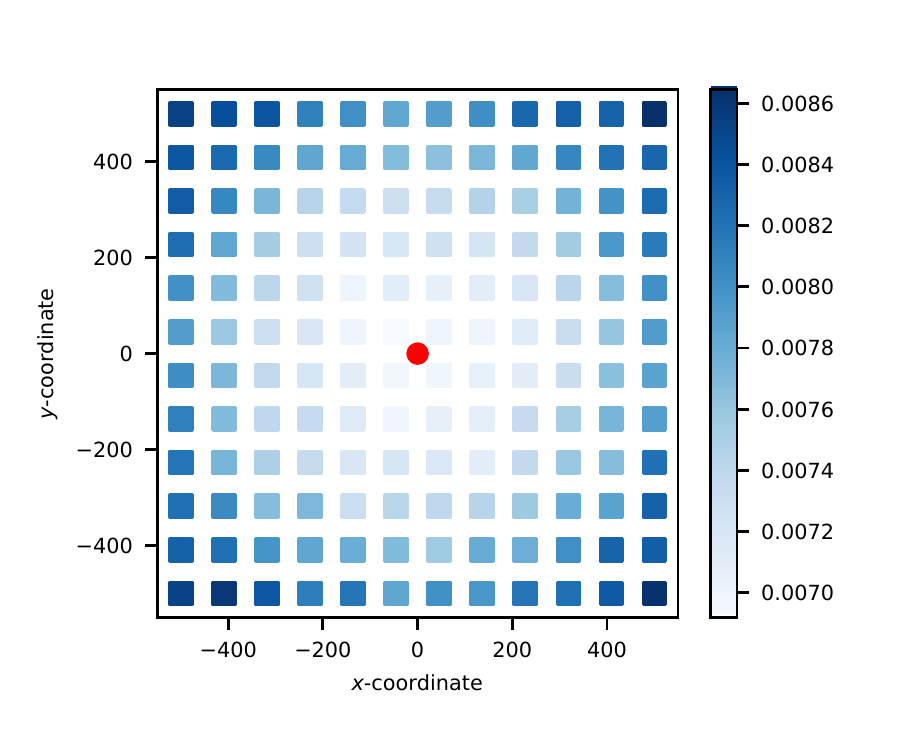}}
\centering
\subfloat[$\vect{p}^*$ in an $M=1$ network with NOMA \newline $~~~~~\text{JF}\big(\bar{R}_1(\vect{p}^*), \cdots,\bar{R}_{144}(\vect{p}^*) \big)=0.9741$]{\includegraphics[width=0.33\linewidth,valign=c]{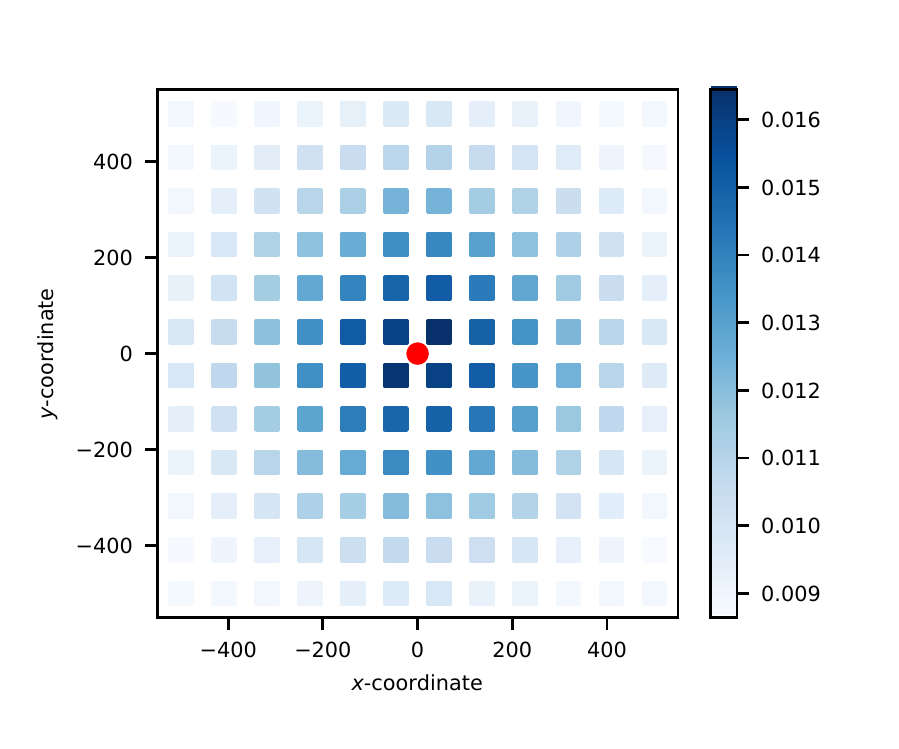}}
\centering
\subfloat[Boxplot of the achievable expected user rates for the considered single- and multi-cell networks with and without NOMA]{\includegraphics[width=0.33\linewidth,valign=c]{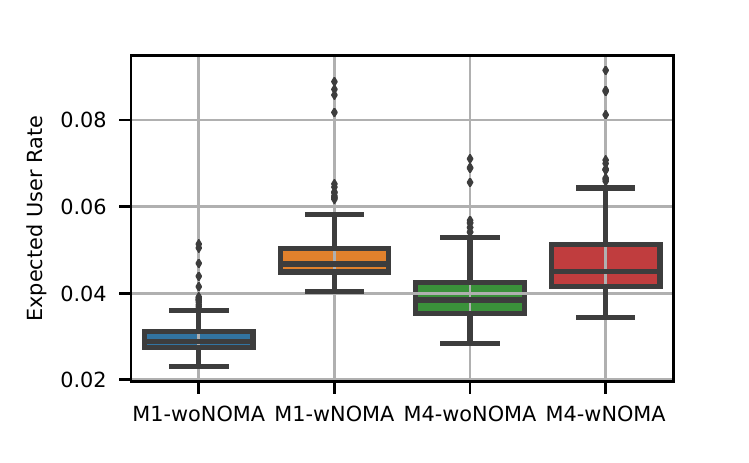}}\\
\subfloat[$\vect{p}^*$ in an $M=4$ network without NOMA \newline $~~~~~~\text{JF}\big(\bar{R}_1(\vect{p}^*), \cdots,\bar{R}_{144}(\vect{p}^*) \big)=0.9657$]{\includegraphics[width=0.33\linewidth,valign=t]{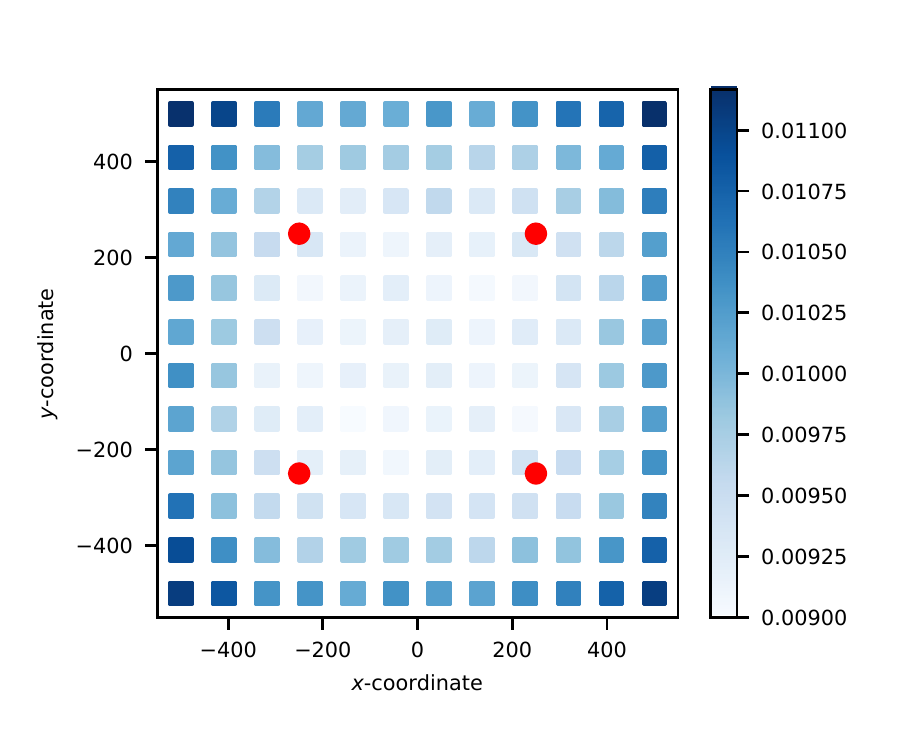}}
\centering
\subfloat[$\vect{p}^*$ in an $M=4$ network with NOMA \newline $~~~~~\text{JF}\big(\bar{R}_1(\vect{p}^*), \cdots,\bar{R}_{144}(\vect{p}^*) \big)=0.9572$]{\includegraphics[width=0.33\linewidth,valign=t]{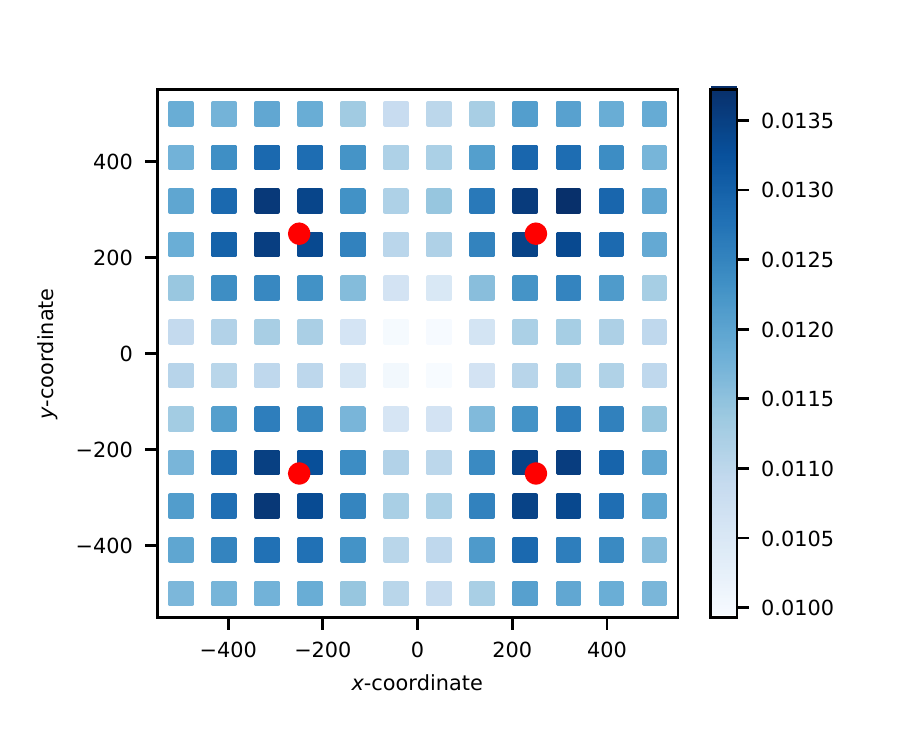}}
\centering
\subfloat[Jain rate fairness gain achieved by optimizing the geometric mean of expected user rates with heterogeneous transmission probabilities. The baseline is Jain rate fairness achieved by optimizing the arithmetic mean of expected user rates with homogeneous transmission probabilities, i.e., $p_i=p, \forall i$.]{\includegraphics[width=0.33\linewidth,valign=t]{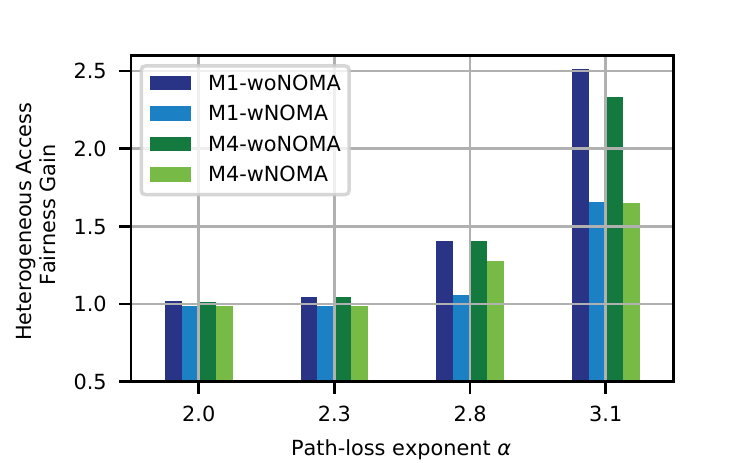}}
\caption{Heatmaps of the optimal heterogeneous access transmission probabilities learned by the distributed algorithm for a mesh grid deployment of $N=144$ IoT devices (blue squares) and (a) $M=1$ BS without NOMA, (b) $M=1$ BS with NOMA, (d) $M=4$ BSs without NOMA, (e) $M=4$ BSs with NOMA. In (a,b,d,e), red circles represent BSs, while the shade of blue squares correspond to the transmission probability of IoT devices (darker shades correspond to higher transmission probability). In (c), a boxplot of the achievable expected user rates for the four networks considered in (a,b,d,e) is shown. In (f), Jain fairness gain of heterogeneous access is shown with respect to optimal homogeneous access that maximizes the arithmetic mean of expected user rates. 
}
\label{fig:hetprob}
\end{figure*}

Last but not least, we look at the transmission probability allocation of users with and without NOMA in a single-cell and a multi-cell IoT network. To this end, we consider networks of $N=144$ IoT devices which are deployed on a mesh grid of $[-500,500] \times [-500,500]$m. In Fig. \ref{fig:hetprob}(a) and Fig. \ref{fig:hetprob}(b), one BS is deployed at $(0,0)$m, whereas in Fig. \ref{fig:hetprob}(d) and Fig. \ref{fig:hetprob}(e) four BSs are deployed at $(-250,-250),~(250,-250),~(-250,250),~(250,250)$m. In Fig. \ref{fig:hetprob}(a) and Fig. \ref{fig:hetprob}(d), heatmaps of the heterogeneous access transmission probabilities learned by the distributed algorithm when the BSs do not support NOMA decoding are shown. As it can be seen from Fig. \ref{fig:hetprob}(a), edge IoT devices have the highest transmission probability (darkest blue shade), while IoT devices which are close to the BS have the lowest transmission probability (lightest blue shade). This allocation strategy maximizes the geometric mean of expected user rates by giving far users more chances to transmit compared with near users, thereby countering the impacts of distance-dependent path-loss on far users in the time domain and achieving a high Jain's rate fairness index of $0.9778$. A similar allocation strategy can be observed in the multi-cell case in Fig. \ref{fig:hetprob}(d), with the notable exception that central users in the middle of the 4 BSs are allocated low transmission probability, although they are relatively far from their respective BSs. This is because these central users cause significant interference to transmissions of users in all cells, unlike users at the edge of the deployment which mainly interfere with transmissions of users from their cell. In this case, the optimal transmission probability allocation also achieves a high Jain's rate fairness index of $0.9657$.

On the other hand, the heatmaps of the heterogeneous access transmission probabilities learned by the distributed algorithm when the BSs support NOMA decoding are shown in Fig. \ref{fig:hetprob}(b) and Fig. \ref{fig:hetprob}(e). It can be seen that the allocation of transmission probabilities is quite different in these cases: users which are far from their serving BS are allocated lower transmission probabilities compared with  users which are close to the BS. Thanks to NOMA decoding, transmissions of near users can be decoded in the first NOMA iteration, while  transmissions of far users are decoded in the second NOMA iteration. Conceptually, a fewer number of near users compete to have their transmissions decoded in the first NOMA iteration, while a larger number of far users compete to have their transmissions decoded in the second NOMA iteration. Hence, near users are allocated higher transmission probability compared with farther users, and NOMA decoding enables a win-win type situation among near-far users, with high Jain's rate fairness indices of $0.9741$ and $0.9572$ for cases (b) and (e), respectively. 

In Fig. \ref{fig:hetprob}(c), a boxplot of the achievable expected user rates for the networks considered in Fig. \ref{fig:hetprob}(a,b,d,e) is shown. The boxplot summarizes the distribution of achievable expected user rates by showing the minimum, the maximum, the median, and the first and third quartiles of the expected user rate, in addition to outliers which are depicted by individual points. It can be observed that when BSs support NOMA decoding, the distribution of expected user rates is shifted upwards which hints that all users benefit from having NOMA-capable BSs. It is also evident that NOMA gain is higher in the single-cell case compared with the multi-cell case, which is consistent with the observations drawn from Fig. \ref{fig:NomaGain}(b). 
Moreover, it can be seen that the variation in expected rates achieved by different users is relatively narrow, and that the are no starving users as all users attain a positive expected rate, which is why the allocation strategies attained by maximizing the geometric mean of expected user rates achieve high Jain's fairness index overall. Note that if the arithmetic mean of expected user rates is to be maximized by choice of $\vect{p}$, the achieved Jain's fairness index will be $\frac{1}{N}$ when BSs do not support NOMA, or at most $\frac{2}{N}$ when BSs support NOMA. This is because the maximum arithmetic mean is attained at one of the extreme points of the convex set $\vect{p} \in [0,1]^N$, which grants the channel exclusively to one or two users without time sharing with other users, leading to very poor fairness. Because maximizing the arithmetic mean by choice of $\vect{p}$ produces adverse solutions, in Fig. \ref{fig:hetprob}(e) we study how our proposed geometric-mean based formulation compares with maximizing the arithmetic mean of expected user rates using a homogeneous transmission probability, that is, $p_i = p, \forall i \in \mathcal{N}$. Fig. \ref{fig:hetprob}(e) shows the heterogeneous access fairness gain, defined as the ratio of Jain rate fairness index achieved by maximizing the geometric mean of expected user rates using our proposed framework, to the Jain rate fairness index achieved by maximizing the arithmetic mean of expected user rates using a homogeneous transmission probability among all users, as a function of the path-loss exponent $\alpha$. It can be observed that tuning the transmission probability of individual users by maximizing the geometric mean yields significant fairness gains up to $250\%$ at higher path-loss exponents. This is because the variance of the received powers at BSs due to the spatial distribution of users is higher when $\alpha$ increases, which necessitates the tuning of the transmission probability at the user level to counter the impacts of distance-dependent path-loss and maintain rate fairness among users. 



\section{Conclusion}
In this paper, we have proposed a novel formulation for random channel access of IoT devices in which the transmission probability of each IoT device is tuned to maximize the geometric mean of users' expected capacity. As the proposed optimization problem is high-dimensional and mathematically intractable, an efficient centralized learning algorithm and a provably convergent distributed learning-based algorithm have been proposed. Our formulation and proposed algorithms provide a versatile {data-driven} framework for optimizing single- and multi-cell random access based wireless IoT networks with and without NOMA decoding techniques, and achieves high capacity fairness among the spatially distributed IoT devices. The proposed framework can be leveraged to evaluate various deployment scenarios and provide guidelines for integrating NOMA techniques in slotted-Aloha systems in support of massive machine type communications in the beyond 5G era.

\section*{Acknowledgment}
This work is supported in part by the NSF grant ECCS1554576 and by the U.S. Department of Energy, Office of Science, under contract DE-AC02-06CH11357. We gratefully acknowledge the computing resources provided on Bebop, a high-performance computing cluster operated by the Laboratory Computing Resource Center at Argonne National Laboratory.

\renewcommand{\thepage} {\arabic{page}}
\bibliographystyle{IEEEtran}
\bibliography{references}
\end{document}